\documentclass[aps,prd,amsmath,amssymb,twocolumn,preprintnumbers,nofootinbib]{revtex4-2}

\usepackage[T1]{fontenc}
\usepackage{times}
\DeclareMathAlphabet{\MATHIT}{OT1}{ptm}{m}{it}
\DeclareSymbolFont{Letters}{OML}{ztmcm}{m}{it}
\DeclareSymbolFontAlphabet{\mathNormal}{Letters}
\usepackage[utf8]{inputenc}
\usepackage[english]{babel}
\usepackage{amsmath}
\usepackage{amsfonts}
\usepackage{amssymb}
\usepackage{amsthm}
\usepackage{enumitem}
\usepackage{dcolumn}
\usepackage{bm}
\usepackage{bbm}
\usepackage{cancel}
\usepackage{tikz}
\usetikzlibrary{calc,shapes,arrows,patterns,decorations.pathmorphing,positioning,automata}
\usepackage{tabularx}
\usepackage{booktabs}
\usepackage{tensor}
\usepackage{scalerel}
\usepackage{xcolor}
\usepackage{mathtools}
\usepackage{float}
\usepackage{xfrac}
\definecolor{darkblue}{rgb}{0,0,.5}
\definecolor{darkgreen}{rgb}{0,0.5,.5}
\definecolor{darkyellow}{rgb}{0.5,0.5,0}
\definecolor{fhl}{rgb}{1,0,0}
\usepackage[pdfborder={0 0 0}]{hyperref}
\hypersetup{colorlinks=true, breaklinks=true, linkcolor=darkblue, menucolor=darkblue, urlcolor=darkblue, linktocpage=true}
\usepackage{geometry}
\usepackage{textpos}
\usepackage{relsize}
\usepackage{balance}
\usepackage[titletoc,title]{appendix}
\allowdisplaybreaks
\usepackage{slashed}
\usepackage{natbib}
\usepackage[export]{adjustbox}
\usepackage[most]{tcolorbox}
\tcbuselibrary{skins}

\newcommand{\MSb}{\overline{\textrm{MS}}}

\newcommand{\e}{\mathrm{e}}
\newcommand{\dd}{\mathrm{d}}

\newcommand{\gGF}{g_{\rm GF}^2}

\newcommand{\gMSb}{g_{\MSb}^2}

\newcommand{\uMSb}{u_{\MSb}}
\newcommand{\uBFM}{u_{\mathrm{BFM}}}
\newcommand{\uGF}{u_{\rm GF}}
\newcommand{\uR}{u_{\rm R}}
\newcommand{\lBFM}{\lambda^{\of{\mathrm{BFM}}}}
\newcommand{\lGF}{\lambda^{\of{\mathrm{GF}}}}

\newcommand{\beq}{\begin{eqnarray}}
\newcommand{\eeq}{\end{eqnarray}}

\newcommand{\bmp}{\noindent\begin{minipage}{16cm}}
\newcommand{\emp}{\end{minipage}\vskip 7mm} 

\def\lsim{\mathrel{\rlap{\lower4pt\hbox{\hskip1pt$\sim$}}
    \raise1pt\hbox{$<$}}}                
\def\gsim{\mathrel{\rlap{\lower4pt\hbox{\hskip1pt$\sim$}}
    \raise1pt\hbox{$>$}}}                

\let\originalleft\left
\let\originalright\right
\renewcommand{\left}{\mathopen{}\mathclose\bgroup\originalleft}
\renewcommand{\right}{\aftergroup\egroup\originalright}
\newcommand{\SU}[1]{\operatorname{SU}\left(#1\right)}
\newcommand{\of}[1]{\left(#1\right)}

\newcommand{\bof}[1]{\biggl(\bigg.#1\bigg.\biggr)}

\newcommand{\sof}[1]{\bigl(\big.#1\big.\bigr)}
\newcommand{\ssof}[1]{(#1)}
\newcommand{\fof}[1]{\left[#1\right]}

\newcommand{\cof}[1]{\left\{#1\right\}}

\newcommand{\avof}[1]{\left\langle #1\right\rangle}

\newcommand{\ssavof}[1]{\small\langle #1\small\rangle}

\renewcommand*\[{\begin{equation}}
\renewcommand*\]{\end{equation}}
\newcommand{\order}{\mathcal{O}}

\newcommand{\upmatrix}[1]{\begin{pmatrix}#1\end{pmatrix}}

\definecolor{newgreen}{RGB}{10,100,20}

\iffalse
\definecolor{refcorrcol}{RGB}{150,0,0}
\newcounter{inrefcorr}
\newcommand{\refcorr}[2][]{\stepcounter{inrefcorr}\ifmmode{{\color{refcorrcol}#2}}\else{{\color{refcorrcol}#2}}\fi\addtocounter{inrefcorr}{-1}}
\everymath{\ifnum\value{inrefcorr}>0\color{refcorrcol}\fi}
\else
\newcommand{\refcorr}[2][]{#2}
\fi

\makeatletter

\renewcommand{\p@paragraph}{}
\makeatother

\begin{document}

\title{Non-perturbative decoupling of massive fermions}

\author{Tobias Rindlisbacher}
\email{trindlis@itp.unibe.ch}
\affiliation{AEC, Institute for Theoretical Physics, University of Bern, Sidlerstrasse 5, CH-3012 Bern, Switzerland}
\author{Kari Rummukainen}
\email{kari.rummukainen@helsinki.fi}
\affiliation{Department of Physics \& Helsinki Institute of Physics,
P.O. Box 64, FI-00014 University of Helsinki}
\author{Ahmed Salami}
\email{ahmed.salami@helsinki.fi}
\affiliation{Department of Physics \& Helsinki Institute of Physics,
P.O. Box 64, FI-00014 University of Helsinki}
\author{Kimmo Tuominen}\email{kimmo.i.tuominen@helsinki.fi}
\affiliation{Department of Physics \& Helsinki Institute of Physics,
P.O. Box 64, FI-00014 University of Helsinki}

\begin{abstract}
SU(2) gauge theory with $N_f=24$ massless fermions is non-interacting at long distances, i.e. it has an infrared fixed point at vanishing coupling.  With massive fermions the fermions are expected to decouple at energy scales below the fermion mass, and the infrared behavior is that of confining SU(2) pure gauge theory.  We demonstrate this behavior non-perturbatively with lattice Monte Carlo simulations by measuring the gradient flow running coupling.
\end{abstract}
\preprint{HIP-2021-27/TH}
\maketitle

\paragraph*{Introduction. --}
Non-Abelian gauge field theories are at the core of the Standard Model of particle physics as well as many of its extensions.  The behavior of these theories is largely dictated by their fermionic matter content.
Due to their applications in beyond Standard Model scenarios, asymptotically free theories with an infrared fixed point \cite{Sannino:2004qp,Hill:2002ap,Dietrich:2005jn,Arbey:2015exa} have recently attracted attention. On the lattice the properties of this type of theories have been studied for SU(2) gauge theory with matter fields
in the fundamental~\cite{Karavirta:2011zg,Leino:2017lpc,Leino:2017hgm,Leino:2018qvq,Amato:2018nvj} or adjoint~\cite{Hietanen:2008mr,Hietanen:2009az,DelDebbio:2008zf,DelDebbio:2009fd,DelDebbio:2010hu,Bursa:2011ru,DeGrand:2011qd,Rantaharju:2015yva,DelDebbio:2015byq} representation.

Much less is known about the dynamics of theories which are not asymptotically free, i.e. where the coupling constant does not vanish at high energies.
For SU($N$) gauge theory with fundamental representation Dirac fermions this happens when the number of fermions $N_f$ is larger than $11N/2$.  While these theories are not directly relevant for the Standard Model, they pose a challenge for our understanding of the gauge field dynamics and the applicability of lattice computation methods.

More concretely, let us consider the evolution of the coupling constant in SU(2) gauge theory with $N_f=24$ Dirac fermions.
If fermions are massless, the theory is non-interacting at long distances, i.e. it has an infrared (IR) fixed point at vanishing coupling \cite{Gross:1973id,Politzer:1973fx}.  At shorter distances (high energy) the coupling is expected to grow until it
diverges at an ultraviolet (UV) Landau pole.
This conclusion is supported by our earlier study of the evolution of the coupling with massless fermions \cite{Leino:2019qwk}.

A non-vanishing fermion mass $m$ introduces an additional scale to the system.  While the UV properties remain to a large extent unaffected by this, the IR physics changes dramatically: fermions are expected to decouple at energy scales $\mu \ll m$ (distance scales $\lambda \gg 1/m$), and the theory behaves like a confining pure gauge SU(2) theory with coupling which grows in the infrared.
Thus, the expectation is that the coupling constant has a minimum near energy scale $\mu \sim m$ ($\lambda \sim 1/m$).  In terms of the renormalization group evolution, the gauge coupling is an irrelevant parameter and the mass a relevant parameter at this minimum.

In this work we measure the evolution of the coupling constant non-perturbatively on the lattice as the fermion mass is varied.  We observe unambiguously the fermion decoupling and the reversal of the coupling constant evolution.  The results agree well with the perturbative predictions in BF-MOM \cite{Rebhan:1985yf,Jegerlehner:1998zg} and massive gradient flow \cite{Harlander:2016vzb,Harlander:2021esn} schemes.  Together with the mass spectrum and scaling laws measured in ref.~\cite{Rantaharju:2021iro}, this gives us a consistent non-perturbative picture of the behavior of the theory from IR to UV scales.

\paragraph*{Perturbative renormalization group evolution. --}

\begin{figure}[htbp]
\centering
\begin{minipage}[t]{0.66\linewidth}
\includegraphics[height=1.04\linewidth,keepaspectratio,margin=3pt 0pt,right]{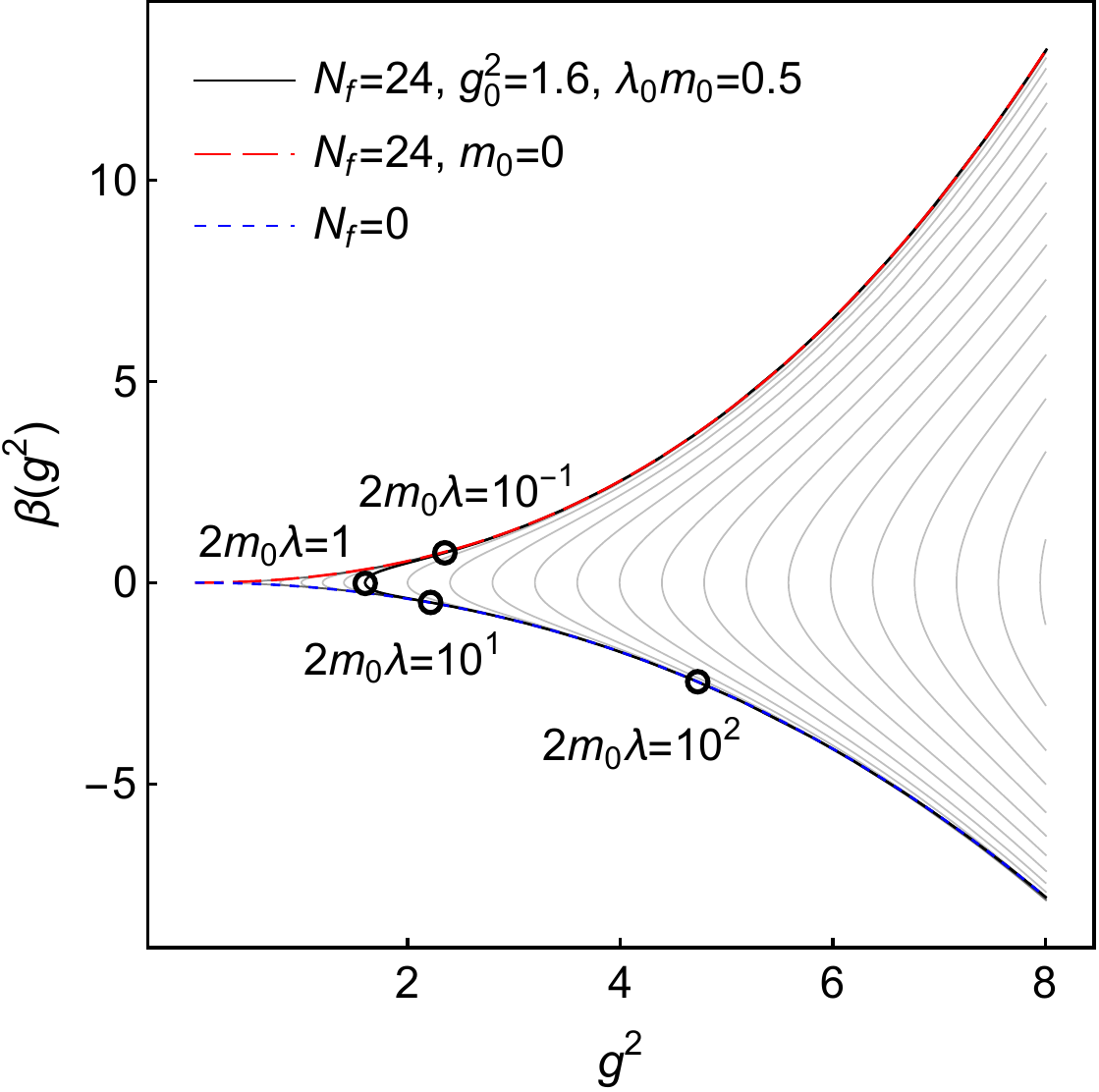}
\end{minipage}\\[5pt]
\begin{minipage}[t]{0.66\linewidth}
\includegraphics[height=1.04\linewidth,keepaspectratio,margin=3pt 0pt,right]{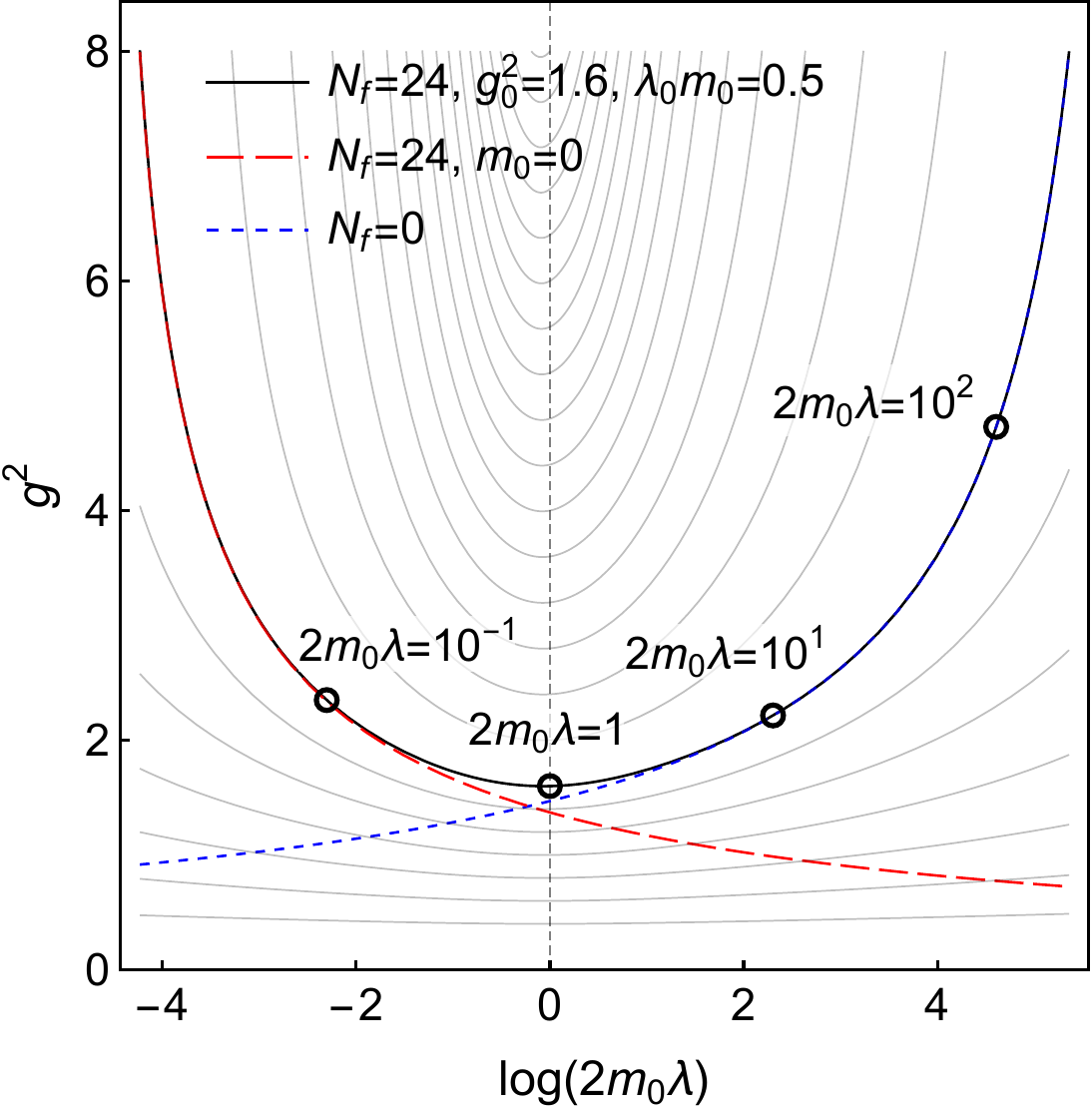}
\end{minipage}\\[-5pt]
\caption{{\em \refcorr{Top:}} 2-loop $\beta$-functions for $N_f=0$ theory (blue, short-dashed) and massless (red, long-dashed) and massive (solid, black) $N_f=24$ theories.  The massive running coupling reaches the minimum value of $g^2 = g_0^2 = 1.6$ at $2 m_0 \lambda_0 \approx 1$.
{\em \refcorr{Bottom:}} couplings $g^2$ as functions of the length scale $\lambda$, obtained by integrating the $\beta$- and $\gamma$-functions.  The integration constants have been set to match the asymptotic behavior.  In both panels we show the evolution when $\lambda$ decreases or increases by a factor of 10. The solid gray lines show the curves for the massive $N_f=24$ theory for different values of $g_0^2$. For $g_0^2>1.6$, $g_0^2$ changes by $\Delta g_0^2=0.4$ between successive curves, whereas for $g_0^2<1.6$, $g_0^2$ changes by $\Delta g_0^2=-0.2$.}
\label{fig:massiverunningcoupling}
\end{figure}

\refcorr{In a mass-dependent renormalization scheme} the evolution of the coupling constant $g^2$ and the mass $m$ \refcorr{are in general determined by a pair of renormalization group (RG) equations}:
\begin{subequations}\label{eq:massivergeq}
\begin{align}
\frac{d g^2}{d \log\of{\lambda}}&=-\beta\ssof{g^2,\lambda\,m}\ ,\label{eq:massivergeqbeta}\\
\frac{d \log\of{m}}{d \log\of{\lambda}}&=\gamma\ssof{g^2,\lambda\,m}\ ,\label{eq:massivergeqgamma}
\end{align}
\end{subequations}
where $\beta$ and $\gamma$ in \eqref{eq:massivergeq} depend on $\lambda\,m = m/\mu$, i.e. on $\lambda$ relative to the scale $1/m$ set by the fermion mass.
In a perturbative expansion the expressions for the mass-dependent $\beta$ and $\gamma$ can be written as
\begin{subequations}\label{eq:massivebetagamma}
\begin{align}
\beta\ssof{g^{2},\lambda\,m}&=-2\,g^2\,\sum\limits_{n=0}^{\infty}\beta_{n}\of{\lambda\,m}\cdot\bof{\frac{g^2}{\of{4 \pi}^2}}^{n+1}\ ,\label{eq:massivebetafunc}\\
\gamma\ssof{g^{2},\lambda\,m}&=\sum\limits_{n=0}^{\infty}\gamma_{n}\of{\lambda\,m}\cdot\bof{\frac{g^2}{\of{4 \pi}^2}}^{n+1}\ .\label{eq:massiveanomdim}
\end{align}
\end{subequations}
\refcorr{We employ the background field momentum subtraction (BF-MOM) scheme in the Landau gauge~\cite{Rebhan:1985yf}. 
We use the 2-loop result from~\cite{Jegerlehner:1998zg}, where the fermion mass dependency in the BF-MOM scheme is implemented in terms of a pole mass~\cite{Hagiwara:1982ct}.
The 2-loop running coupling can therefore be determined form the beta-function alone, with the first two coefficients given by:
\begin{subequations}\label{eq:bfmombetas}
\begin{align}
\beta_{0}\of{\lambda\,m_0}&=\frac{11}{3}\,C_{G} - \frac{4}{3}\,T_R\,N_f\,b_{0}\of{x}\ ,\label{eq:bfmombeta0}\\
\beta_{1}\of{\lambda\,m_0}&=\frac{34}{3}\,C^2_{G} - T_R\,N_f\,b_{1}\of{x}\ ,\label{eq:bfmombeta1}
\end{align}
\end{subequations}
with $x=-1/\of{2\,\lambda\,m_0}^{2}$ and $m_0$
the fermion pole mass}. For SU(2) gauge theory $C_G=2$ and
for fundamental representation fermions $T_R = 1/2$.
The expressions for the coefficients $b_0\of{x}$ and $b_1\of{x}$ can be found in \cite{Jegerlehner:1998zg}.

In Fig.~\ref{fig:massiverunningcoupling} we illustrate the \refcorr{behavior of the 2-loop massive $\beta$-function as function of the running coupling (top) and the running coupling itself as a function of the length scale (bottom)}.
The evolution curves are obtained by integrating \refcorr{Eq.~\eqref{eq:massivergeqbeta} numerically (cf. Appendix~\ref{par:numint}) with \eqref{eq:bfmombetas}}, starting from initial values $(g^2,\lambda)=(g_0^2,\lambda_0)$. \refcorr{As Eq.~\eqref{eq:massivergeq} depends merely on the product of fermion pole mass $m_0$ and length scale $\lambda$, we set $\lambda_0=1/(2\,m_0)$. The different trajectories correspond to different choices of $g_0^2$}.

We note that the $\beta$-functions have zeroes at $2\lambda m_0 \approx 1$.  These correspond to local minima of the coupling, not to fixed points, because $\beta(g^2,\lambda m_0)$ is vertical here.
Our choice of the initial value $g_0^2$ is very close to the minimum value of the coupling along the evolution curve.

As an example, the evolution curves with $g_0^2 = 1.6$ are highlighted in Fig.~\ref{fig:massiverunningcoupling}.
For comparison, the figures also show the corresponding asymptotic cases of $N_f=24$ massless fermions, and pure gauge (infinitely heavy and therefore completely decoupled fermions).

\paragraph*{Lattice setup. --}

We simulate a $\SU{2}$ gauge theory coupled to $N_{f}=24$ mass-degenerate, dynamical fermions in the fundamental representation. The corresponding lattice action can be summarized as follows:
\begin{displaymath}
S = S_G(U) + S_F(V) + c_{\rm SW}\,S_{\rm SW}(V)\ ,\label{eq:lataction}
\end{displaymath}%
where $U$ represents $\SU{2}$ gauge link matrix in the fundamental representation, $V$ the corresponding hypercubically truncated stout smeared link matrix (HEX smearing)~\cite{Capitani:2006ni}, $S_G$ is the Wilson gauge action, and $S_F$ and $S_{\rm SW}$ are, respectively, the Wilson fermion action and clover term with Sheikholeslami-Wohlert coefficient $c_{\rm SW} = 1$ \cite{Rantaharju:2015yva}.

Simulations are carried out using a hybrid Monte Carlo (HMC) algorithm with leapfrog integrator and chronological initial values for the fermion matrix inversion~\cite{Brower:1995vx}.
The HMC trajectories have unit-length and the number of leapfrog steps is set to yield acceptance rates above 80\%.

The lattice quark mass is determined via PCAC relation:
\begin{equation}
 a\,m_q = \left.\frac{(\partial_4^\ast+\partial_4)f_A(x_4)}{4\,f_P(x_4)}\right\vert_{x_4 = L/2} \ ,
 \label{eq:mq}
\end{equation}
where $a$ is the lattice spacing, $\partial_4$ and $\partial_4^{*}$ are forward and backward lattice time-derivative operators, and $f_A$ and $f_P$ are axial and pseudoscalar current correlation functions \cite{Luscher:1996ug}. Eq.~\eqref{eq:mq} receives an $O(a)$ correction, but for smeared quarks it is very small and we omit it here.  The bare quark mass from \eqref{eq:mq} is multiplicatively renormalized;  however, the renormalization is expected to vary little as the coupling is changed, see e.g. \cite{Campos:2018ahf}, and for our purposes the bare mass is sufficient.

\begin{figure*}[htbp]
\centering
{\begin{minipage}[t]{0.33\linewidth}
\includegraphics[height=1.04\linewidth,keepaspectratio,margin=3pt 0pt,right]{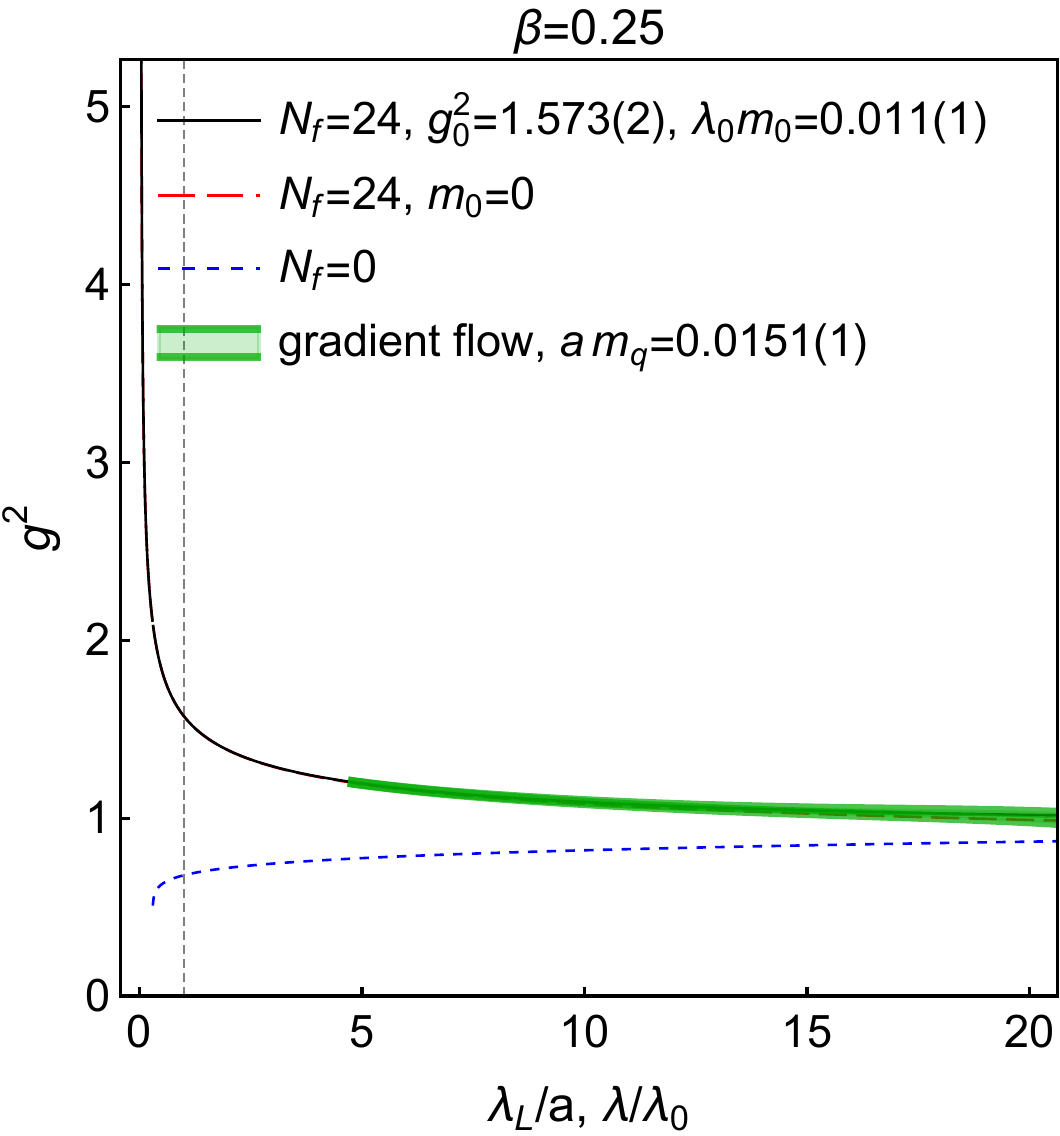}
\end{minipage}\hfill
\begin{minipage}[t]{0.33\linewidth}
\includegraphics[height=1.04\linewidth,keepaspectratio,margin=3pt 0pt,right]{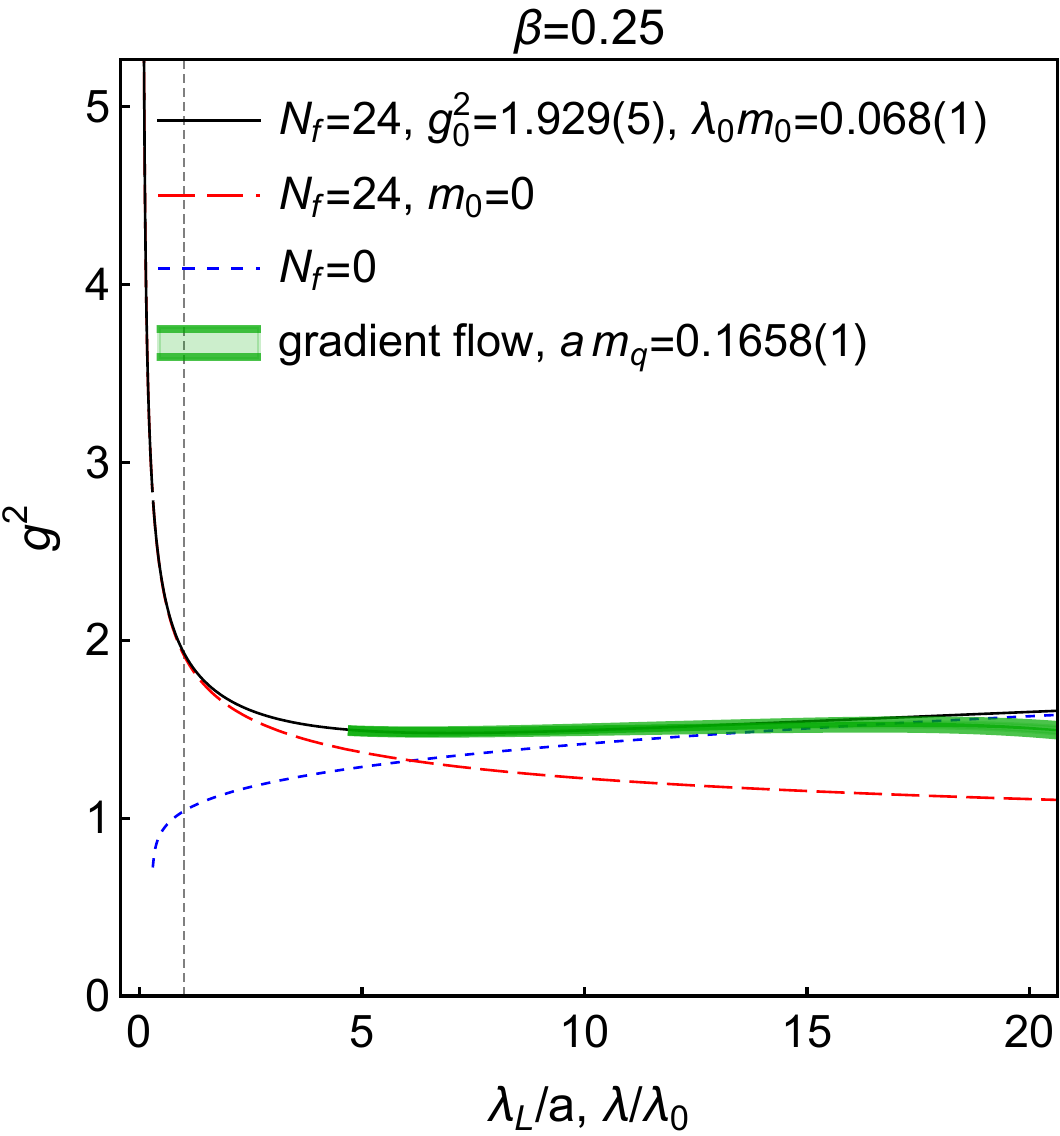}
\end{minipage}\hfill
\begin{minipage}[t]{0.33\linewidth}
\includegraphics[height=1.04\linewidth,keepaspectratio,margin=3pt 0pt,right]{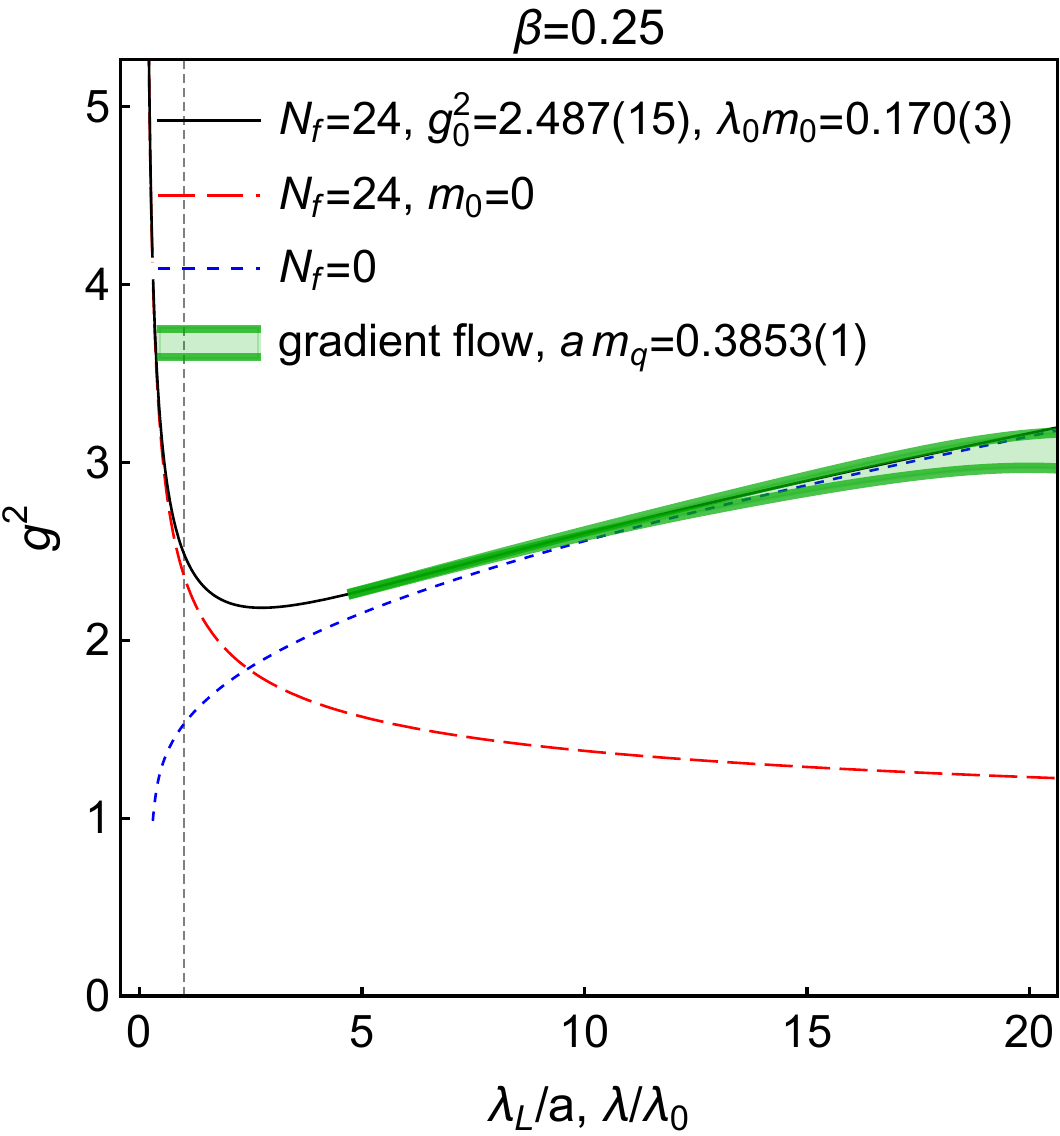}
\end{minipage}}\\[3pt]
{\begin{minipage}[t]{0.33\linewidth}
\includegraphics[height=1.04\linewidth,keepaspectratio,margin=3pt 0pt,right]{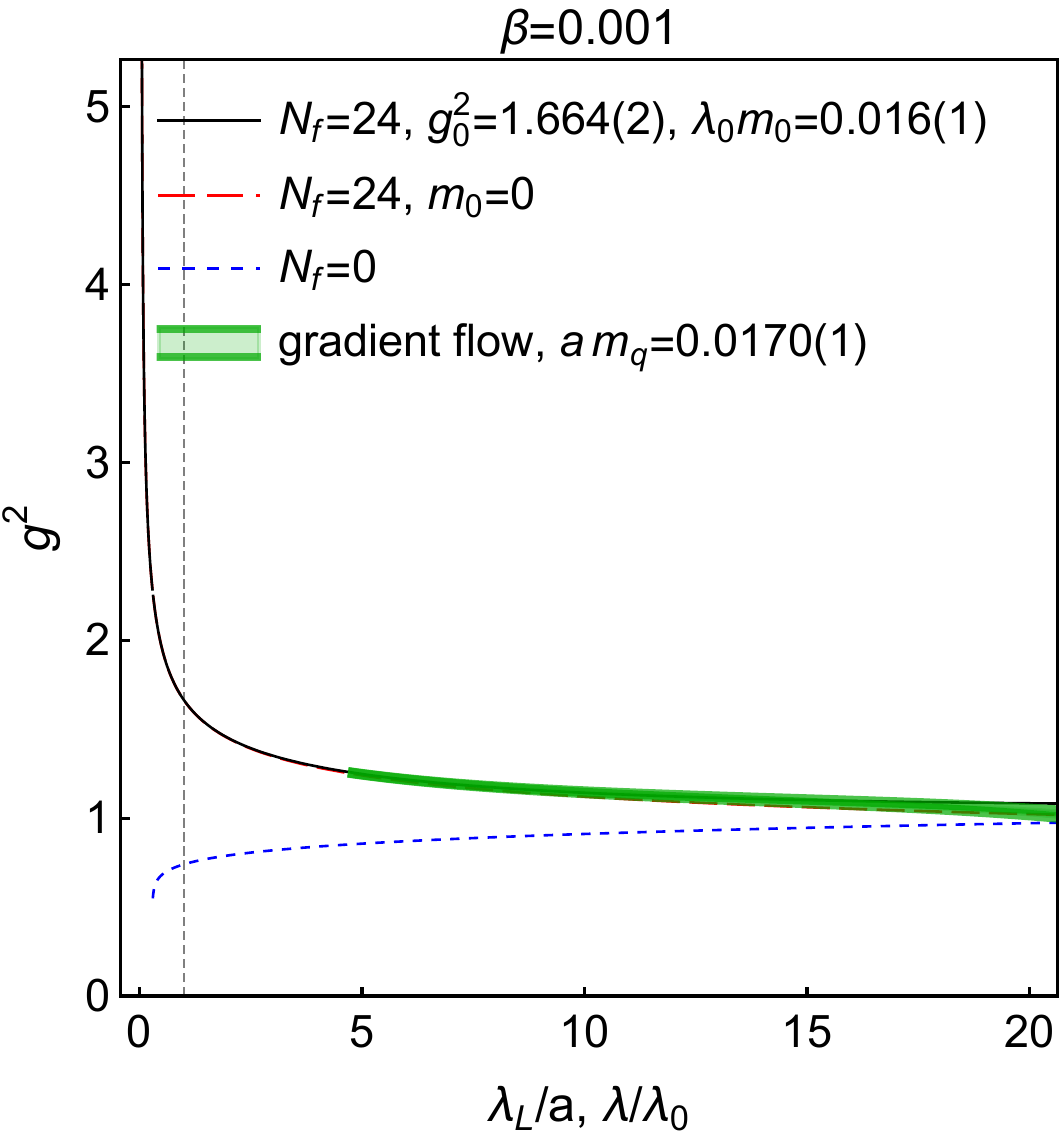}
\end{minipage}\hfill
\begin{minipage}[t]{0.33\linewidth}
\includegraphics[height=1.04\linewidth,keepaspectratio,margin=3pt 0pt,right]{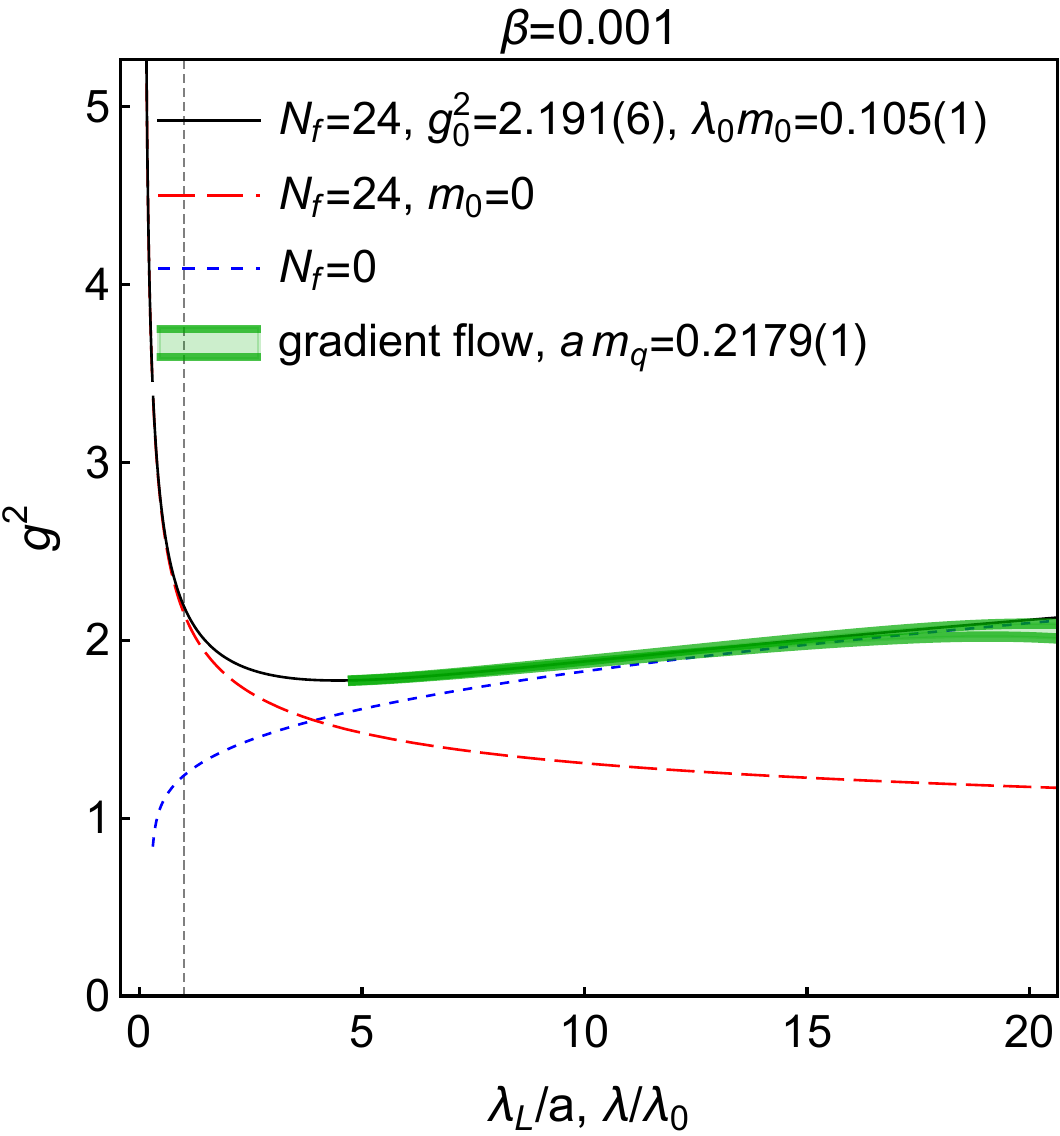}
\end{minipage}\hfill
\begin{minipage}[t]{0.33\linewidth}
\includegraphics[height=1.04\linewidth,keepaspectratio,margin=3pt 0pt,right]{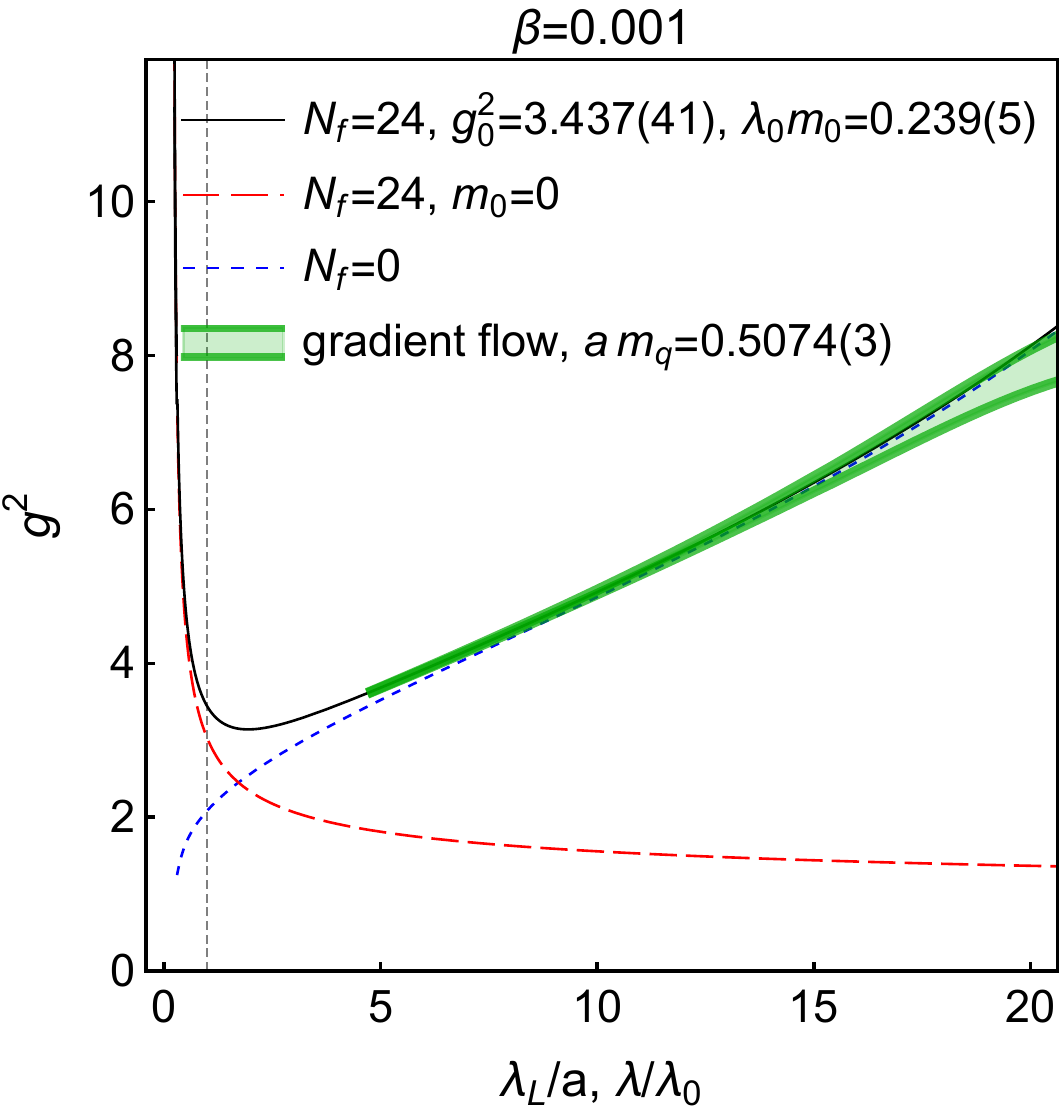}
\end{minipage}}\\[-5pt]
\caption{The measured gradient flow running coupling (green bands), obtained on a $V=(48a)^4$ lattice at $\beta\in\cof{0.25,0.001}$ at three quark masses $m_q$, to which two-loop running coupling is fitted (solid black line). The gradient flow length scale $\lambda_L$ is shown in interval $\lambda_L/a \in [4.8,20]$.
In comparison, the matched pure gauge SU(2) (blue dotted line) and $N_f=24$, $m_q=0$ (red dashed line) couplings are also shown.}
\label{fig:couplingvsscalefit}
\end{figure*}

We use the running coupling in the non-perturbative gradient flow (GF) scheme~\cite{Luscher:2010iy,Fodor:2017die,Peterson:2021lvb}. In the continuum, it can be written as function of length scale $\lambda$ as~\cite{Luscher:2010iy}:
\[
\gGF\of{\lambda}=\frac{2\,\pi^2\,\lambda^4\,\avof{E\of{\lambda}}}{3\of{N^2-1}}\ ,\label{eq:gsqgf}
\]
with $\avof{E\of{\lambda}}$ being the flow-evolved gauge action at flow time $t=\lambda^2/8$.
On a lattice of size $L^4$ with periodic boundary conditions for the gauge field, we use \refcorr{
\[
\gGF\of{\lambda_L,L}=\frac{2\,\pi^2\,\lambda_L^4\,\avof{E\of{\lambda_L,L}}}{3\,\of{N^2-1}\,\of{1+\delta_{L/a}\of{\lambda_{L}/L}}}\ ,\label{eq:latgsq}
\]}
as an estimator for Eq.~\eqref{eq:gsqgf}, with lattice flow scale $\lambda_L$. Here $\avof{E\of{\lambda_L,L}}$ is the expectation value of the clover energy of the flow-evolved lattice gauge field after flow time $t=\lambda_L^2/8$, and\refcorr{
\[
\delta_N\of{c}=\bof{\sqrt{\pi} c\,\sum\limits_{\mathclap{n=-N/2}}^{\mathclap{N/2-1}}\,\e^{-\of{N c \sin\of{\pi n/N}}^2}}^4-\frac{\pi^2 c^4}{3}-1\,\ \label{eq:finitevolcorr}
\]}
is a finite volume \refcorr{and finite lattice spacing} correction for $\ssavof{E\of{\lambda_L,L}}$. Eq.~\eqref{eq:finitevolcorr} is obtained from the corresponding expression in~\cite{Fodor:2012td} by replacing continuum with lattice momenta (cf.~\cite{Fritzsch:2013je}). The flow is governed by the Lüscher-Weisz action~\cite{Luscher:1984xn}.

To relate the GF scheme from Eq.~\eqref{eq:gsqgf} to the BF-MOM scheme from Fig.~\ref{fig:massiverunningcoupling}, we make use of the quark mass dependent one-loop expression for $\ssavof{E\of{\lambda}}$ from \cite{Harlander:2016vzb} to derive~\cite{Harlander:2021esn} the leading coefficient of a perturbative expansion of the mass-dependent GF scheme beta function:
\[
\beta_{0,\mathrm{GF}}\of{\gGF,\lambda\,m}\,=\,\beta_{0}+\frac{4}{3} T_R\,N_f\,x\,\frac{\dd\Omega_{1q}\of{x}}{\dd x}\ ,\label{eq:oneloopgfbeta}
\]
where $\beta_{0}$ is the (universal) leading coefficient of the massless $\MSb$ scheme, $x=-1/\ssof{2 m\lambda}^2$, and $\Omega_{1q}$ is given in~\cite{Harlander:2016vzb}, or with our conventions in Appendix~\ref{par:gfsbf}.
As the renormalization group equations~\eqref{eq:massivergeq} do not fix the overall scale in either scheme, their respective scales $\lambda_{\mathrm{GF}}$ and $\lambda_{\mathrm{BFM}}$ should be related by a rescaling of the form $\lambda_{\mathrm{GF}}=\rho_s\,\lambda_{\mathrm{BFM}}$. To determine $\rho_s$, we require that for a given quark mass, the behavior of the running coupling near the corresponding decoupling point (where the beta function changes sign) is as similar as possible (cf. Appendix~\ref{par:gfvsbfm}) in the two schemes. At the one-loop level, this leads to the matching criterion
\[
\beta_{0,\mathrm{BFM}}\of{\lambda_0\,m_0}=\beta_{0,\mathrm{GF}}\of{\lambda_0\,m_0\,\rho_s}\ ,\label{eq:oneloopbetaequal}
\]
where $\beta_{0,\mathrm{BFM}}$ is the leading BF-MOM beta function coefficient from Eq.~\eqref{eq:massivebetagamma}, and $\lambda_0=1/\of{2\,m_0}$ is the approximate decoupling scale in the BF-MOM scheme. This yields:
\[
\rho_s=2.5359\ldots\ .\label{eq:rescalefac}
\]

\paragraph*{Results. --}
Most of our analysis is done using ensembles of lattices of size $V=L^4$, $L=48a$, with smaller lattices used for finite size analysis. These ensembles were used for the spectrum analysis in~\cite{Rantaharju:2021iro}.  The bare lattice gauge coupling is parametrized with $\beta = 4/g_{0,\text{lat}}^2$, and we use values $\beta = 0.25$, $0.001$ and $-0.25$.  Because Wilson fermions induce a positive shift in effective $\beta$ \cite{Hasenfratz:1993az,Blum:1994xb},
very small and even negative values of $\beta$ are needed to compensate for this effect with large numbers of fermions.  The simulation parameters and the measured PCAC quark masses are listed in Table \ref{tbl:lattices}.

\begin{table}[tb]
\centering
\begin{tabular}{l l c l l l l l}
\hline
\multicolumn{1}{c}{$\beta$} & \multicolumn{1}{c}{$\kappa$}
        & $am_q$ & \multicolumn{1}{c}{$g_0^2$} & \multicolumn{1}{c}{$\lambda_0 m_0$} & \multicolumn{1}{c}{$r_s$} & acc. & stat. \\
  \hline
  -0.25 & 0.1309 & \text{0.0202(1)} & \text{1.783(2)} & \text{0.010(1)} & 0.50 & 0.91 & \text{1.3k} \\
  -0.25 & 0.129 & \text{0.1001(1)} & \text{2.002(4)} & \text{0.052(1)} & 0.52 & 0.91 & \text{2.9k} \\
  -0.25 & 0.1277 & \text{0.1608(1)} & \text{2.192(6)} & \text{0.083(1)} & 0.52 & 0.9 & \text{3.0k} \\
  -0.25 & 0.1263 & \text{0.2266(1)} & \text{2.47(2)} & \text{0.111(2)} & 0.49 & 0.91 & \text{3.2k} \\
  -0.25 & 0.125 & \text{0.3013(2)} & \text{2.74(3)} & \text{0.156(3)} & 0.52 & 0.91 & \text{3.3k} \\
  -0.25 & 0.123 & \text{0.4546(3)} & \text{4.05(4)} & \text{0.203(3)} & 0.45 & 0.91 & \text{3.5k} \\
  ~0.001 & 0.1299 & \text{0.0170(1)} & \text{1.664(2)} & \text{0.016(1)} & 0.94 & 0.92 & \text{2.1k} \\
  ~0.001 & 0.129 & \text{0.0517(1)} & \text{1.760(2)} & \text{0.022(1)} & 0.43 & 0.93 & \text{1.3k} \\
  ~0.001 & 0.125 & \text{0.2179(1)} & \text{2.191(6)} & \text{0.105(1)} & 0.48 & 0.92 & \text{3.2k} \\
  ~0.001 & 0.12 & \text{0.5074(3)} &\text{3.44(5)} & \text{0.239(5)} & 0.47 & 0.93 & \text{3.6k} \\
  ~0.25 & 0.129 & \text{0.0151(1)} & \text{1.573(2)} & \text{0.011(1)} & 0.73 & 0.91 & \text{1.9k} \\
  ~0.25 & 0.125 & \text{0.1658(1)} & \text{1.929(5)} & \text{0.068(1)} & 0.41 & 0.93 & \text{2.9k} \\
  ~0.25 & 0.12 & \text{0.3853(1)} & \text{2.49(2)} & \text{0.170(3)} & 0.44 & 0.93 & \text{3.6k} \\
  ~0.25 & 0.115 & \text{0.7534(7)} & \multicolumn{1}{c}{-} & \multicolumn{1}{c}{-} & \multicolumn{1}{c}{-} & 0.94 & \text{3.4k} \\
  \hline
\end{tabular}\\[-5pt]
\caption{Simulation parameters, PCAC quark mass, fitted $g_0^2$ and $m_0$ (where fit was possible), ratio $r_s=\lambda_0 m_0/\ssof{a m_q}$, HMC acceptance and the total number of gauge configurations used in the analysis. \refcorr{The indicated uncertainties are statistical ones.}}
\label{tbl:lattices}
\end{table}

In Fig.~\ref{fig:couplingvsscalefit} we show examples of $\gGF(\lambda_L)$ for $\beta = {0.25,0.001}$ and three different values of $m_q$, each. The switch from the light quark behavior (left column), where the coupling decreases with distance, to the heavy quark behavior (right column) with increasing coupling is evident.
We compare with the 2-loop perturbation theory by fitting $g_0^2$ and $\lambda_0 m_0$ to match $\gGF$. In effect, the fit procedure achieves the relative multiplicative renormalization between $\lambda m_0$ in BF-MOM scheme and $\lambda_L m_q$ on the lattice.

For concreteness, we set $\lambda_0 = a$ so that $\lambda = \lambda_L$, and determine the pair $(g_0^2,m_0^2)$ corresponding to a given pair $(\beta,m_q)$ by a least squares fit. The fit is carried out over the range $\lambda_L/a \in [4.8,20]$. Thus, on a given lattice, we are able to follow the evolution of the coupling over a scale factor of four.  The fit between the lattice data and the perturbative coupling is in general very good, well within the statistical error of each $\gGF$-curve, with the exception of the largest couplings $\gGF \gsim 10$.  The fit parameters are listed in Table~\ref{tbl:lattices}.

\begin{figure}[tb]
\centering
\includegraphics[width=0.7\linewidth,keepaspectratio]{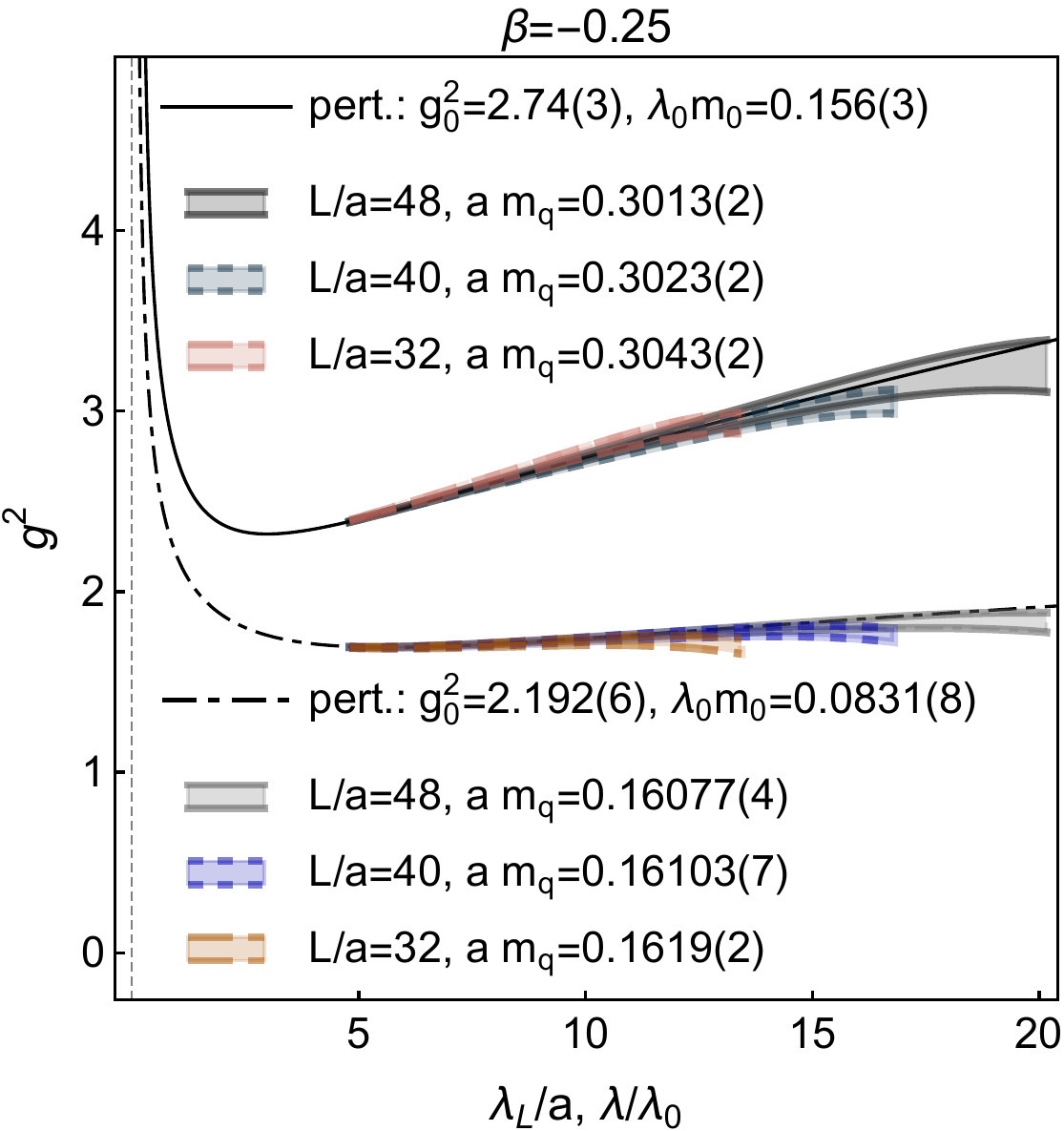}\\[-5pt]
\caption{$\gGF$ measured from $L=32a$, $40a$ and $48a$ lattices at $\beta=-0.25$, $\kappa=0.125$ ($a m_q\approx 0.30$) resp. $\kappa=0.1277$ ($a m_q\approx 0.16$). The flow is plotted in the interval $4.8a < \lambda_L < 0.42L$.}
\label{fig:finiteV}
\end{figure}

\begin{figure}[tb]
\centering
\includegraphics[width=0.8\linewidth,keepaspectratio]{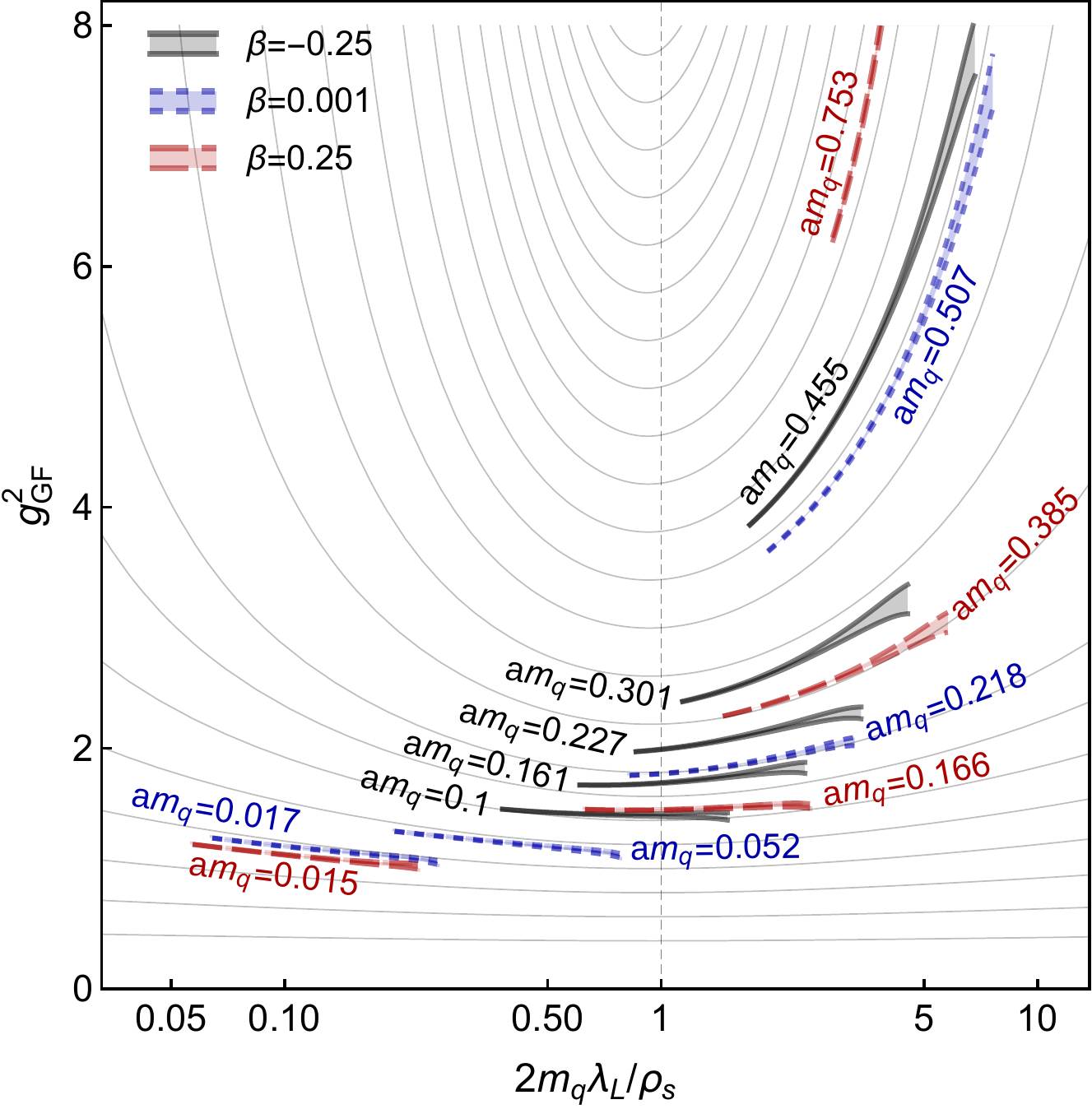}\\[-5pt]
\caption{The \refcorr{lattice gradient flow running coupling} (black, blue and red error bands) as function of $2\,m_q\,\lambda_L/\rho_s$, where $m_q$ is the PCAC quark mass and $\rho_s$ is given by Eq.~\ref{eq:rescalefac}. The data is superimposed to the massive 2-loop gradient flow curves from the \refcorr{lower} panel of Fig.~\ref{fig:massiverunningcoupling}.}
\label{fig:couplingvsmassnormedscale}
\end{figure}

We have checked the finite volume effects by analyzing $\gGF$ on lattices with $L/a = 32$, $40$ and $48$.
The volume dependence is small \refcorr{and for $\lambda_L/L\in\fof{0.1,0.4}$} within the statistical errors\refcorr[with only weak systematic behavior observed]{}; an example of this is shown in Fig.~\ref{fig:finiteV}.  Thus, we do the final analysis using only the largest volume results.

From Table~\ref{tbl:lattices} we can observe that $am_q\,r_s \sim \lambda_0 m_0$ with \refcorr{$r_s\approx 0.4-0.5$}{} in all cases where the fit is reliable (at very small $m_q$ the fit becomes compatible with a vanishing quark mass). This \refcorr[implies]{suggests} that the relative scale renormalization between the lattice and the BF-MOM scheme is approximately constant in the studied range and the rescaling factor $r_s=1/\rho_s$ \refcorr{can be considered compatible with Eq.~\eqref{eq:rescalefac} with the given systematic uncertainties}{}.

In Fig.~\ref{fig:couplingvsmassnormedscale} we plot all measurements of $\gGF$ against $2m_q\lambda_L/\rho_s$, with $\rho_s$ from Eq.~\eqref{eq:rescalefac}, overlaid with the perturbative $g^2$ from Fig.~\ref{fig:massiverunningcoupling}. There are no fitted parameters. The lattice data follows the 2-loop perturbative curves remarkably well, independent of the value of $\beta$. \refcorr{We have verified that the 2-loop BF-MOM running coupling should be a good approximation to the perturbative massive RG running coupling~(cf. Appendix \ref{par:gfvsbfm}-\ref{par:3lnonuniv}).}
There are cases where simulation results with different $\beta$ and $m_q$ fall on curves which are very close to each other. Since different values of $\beta$ \refcorr{and $m_q$ imply in general} different lattice spacings, this demonstrates that the results scale when the lattice spacing is varied. In contrast to the asymptotically free lattice QCD, the lattice spacing becomes smaller when $\beta$ is decreased, and the theory does not have a continuum limit because of the UV Landau pole.

\paragraph*{Conclusions. --}
Using SU(2) gauge theory with $N_f=24$ fermions of mass $m$ we have presented a clear non-perturbative demonstration of the decoupling of fermions at distance scale $\sim 1/m$.  At the same time, the behavior of the theory changes dramatically:
at smaller distances the theory behaves as IR trivial, non-asymptotically free theory, with coupling decreasing with distance, whereas at longer distances it behaves like pure gauge SU(2) theory with increasing coupling.
Together with the study of the excitation spectrum in this theory \cite{Rantaharju:2021iro}, this provides a consistent non-perturbative description of the behavior of the theory from IR to UV scales.

\paragraph*{Acknowledgment. --}
\label{Aknwldg}
The support of the Academy of Finland grants 308791, 310130 and 320123 is acknowledged. The authors wish to acknowledge CSC - IT Center for Science, Finland, for generous computational resources.

\section*{Appendix}
This appendix provides technical details on (\ref{par:numint}) the numerical method used to integrate the RG equations, (\ref{par:gfsbf}) the method used to obtain the gradient flow beta function, and (\ref{par:gfvsbfm}) the method used to match the gradient flow and BF-MOM schemes, as well as (\ref{par:1lvs2lbfmbf}) a comparison of the one- and two-loop perturbative running couplings, and (\ref{par:3lnonuniv}) a discussion of possible effects from 3-loop non-universality in the massless and pure gauge limit.\\[\parsep]

\paragraph{Numerical integration of RG equations. --}\label{par:numint}
While the first-order differential equations,
\begin{subequations}\label{eq:massivergeq}
\begin{align}
\frac{d u}{d \log\of{\lambda/\lambda_0}}&=-\beta\ssof{u,x}\ ,\\
\frac{d \log\of{m}}{d \log\of{\lambda/\lambda_0}}&=\gamma\ssof{u,x}\ ,
\end{align}
\end{subequations}
where $x=-1/\ssof{2\,\lambda\,m}^2$, which describe the running of the coupling, $u=g^2$, and of the quark mass $m$ as the length-scale $\lambda$ is changed ($\lambda_0$ is an arbitrary reference scale), can straightforwardly be integrated with some Runge-Kutta method, we will here briefly discuss an alternative integration approach, which might be useful when using a mathematics program which is primarily meant for symbolic evaluation and whose built-in differential equation solvers have difficulties if the functions appearing in the differential equations need to be evaluated with high numerical accuracy. In the case of the $\SU{2}$ gauge theory with $N_f=24$ massive flavors in the BF-MOM scheme~\cite{Rebhan:1985yf,Jegerlehner:1998zg}, employed in the main text, this can in particular happen for RG-trajectories that pass through small coupling values, {\refcorr{where the running becomes very slow}.

We first note, that for any function $F\of{u,x}$, which depends on the coupling $u\of{\lambda}=g^2\of{\lambda}$ and on the variable $x=-1/\ssof{2\,\lambda\,m}^2$, where $m\of{\lambda}$ is the quark mass, one can write the total derivative with respect to the log of the scale ratio $\lambda/\lambda_0$ as:
\begin{multline}
\frac{\dd F\of{u,x}}{\dd\log\of{\lambda/\lambda_0}}=\underbrace{\lambda\frac{\dd u}{\dd\lambda}}_{\mathclap{-\beta\of{u,x}}}\frac{\partial F}{\partial u}+\bof{\lambda\frac{\partial x}{\partial \lambda}+\underbrace{\lambda\frac{\dd m}{\dd\lambda}}_{\mathclap{m\,\gamma\of{u,x}}}\frac{\partial x}{\partial m}}\frac{\partial F}{\partial x}\\
=-\beta\of{u,x}\frac{\partial F}{\partial u}+\bof{\underbrace{\lambda\frac{\partial x}{\partial \lambda}}_{-2\,x}+\gamma\of{u,x}\,\underbrace{m\frac{\partial x}{\partial m}}_{-2\,x}}\frac{\partial F}{\partial x}\\
=-\beta\of{u,x}\frac{\partial F}{\partial u}-2\,x\of{1+\gamma\of{u,x}}\frac{\partial F}{\partial x}\ ,
\end{multline}
where we used the definitions of the beta and gamma functions from Eq.~\eqref{eq:massivergeq} to get from the first to the second line.

We can therefore easily Taylor expand the running coupling, $u$, and the running mass, $m$, around some reference scale $\lambda_0$:
\begin{multline}
\upmatrix{u\of{\lambda}\\ \log\of{m\of{\lambda}/m_0}}=\\
\upmatrix{u_0\\ 0}+\sum_{n=1}^{\infty}\,c_n\of{u_0,x_0}\,\frac{\log^n\of{\lambda/\lambda_0}}{n!}\ ,\label{eq:taylorexpandedflow}
\end{multline}
where we used the shorthand notation $u_0=u\ssof{\lambda_0}$, $m_0=m\ssof{\lambda_0}$ and $x_0=x\ssof{\lambda_0}$, and the expansion coefficients, $c_n$, are determined by setting
\[
c_1\of{u,x}=\upmatrix{-\beta\of{u,x}\\\gamma\of{u,x}}\label{eq:cinit}
\]
and using the recurrence relation
\begin{multline}
c_{n+1}\of{u,x}=-\beta\of{u,x}\frac{\partial c_n\of{u,x}}{\partial u}\\
-2\,x\of{1+\gamma\of{u,x}}\frac{\partial c_n\of{u,x}}{\partial x}\ .\label{eq:citer}
\end{multline}

To integrate the RG flow, we will, however, not resort to the full Taylor expansion, but keep only the terms up to e.g. $n=3$. We can then determine a maximum log-distance of scales $|\log\ssof{\lambda/\lambda_0}|$ for which the truncation error is still negligible and use the corresponding $\lambda$-value to update:
\[
u_0=u\of{\lambda} ,\ m_0=m\of{\lambda} ,\ x_0=x\of{\lambda} ,\ \lambda_0=\lambda\ ,
\]
and repeat the whole procedure with the so-obtained new values for $u_0$, $m_0$, and $\lambda_0$, and so on. The integration can in this way be carried out in both direction, i.e. towards increasing and decreasing scales.

\refcorr{If the mass parameter used in Eqs.~\eqref{eq:massivergeq} is not a running mass but e.g. a pole mass, we can simply set $\gamma\of{u,x}=0$ 
for all $u$ and $x$, and drop the mass-components from all terms in Eqs.~\eqref{eq:taylorexpandedflow}-\eqref{eq:citer}.}

\paragraph{Gradient flow scheme beta function. --}\label{par:gfsbf}
A beta function for the gradient flow running coupling in the presence of a non-zero quark mass can be defined by using the expression given in~\cite{Harlander:2016vzb} for the leading quark mass corrections to the gradient flow-evolved gauge field energy, $\avof{E\of{t}}$, in the (massless) $\MSb$ scheme. By defining the gradient flow running coupling to be given in terms of the flow-evolved gauge-field energy via~\cite{Luscher:2010iy}
\[
\gGF\of{t}:=\frac{128\,\pi^2\,t^2\,\avof{E\of{t}}}{3\,\of{N^2-1}}\
\]
one then finds with~\cite{Harlander:2016vzb}, that
\begin{multline}\label{eq:gfrunningcoupling}
\uGF\of{\lambda}=\uMSb\of{\lambda_0}\bof{1+k_1\of{\lambda/\lambda_0}\frac{\uMSb\of{\lambda_0}}{\of{4\pi}^2}\\
+l_1\of{\lambda\,m\of{\lambda_0}}\frac{\uMSb\of{\lambda_0}}{\of{4\pi}^2}+\order\of{\uMSb^2\of{\lambda_0}}}\ ,
\end{multline}
with $\uGF=\gGF$, resp. $\uMSb=\gMSb$, $\lambda=\sqrt{8\,t}$ being the flow scale corresponding to flow-time $t$, and $\lambda_0$ is the renormalization (length-)scale at which $\uMSb\of{\lambda_0}$ and the quark mass $m\of{\lambda_0}$ are defined.
The quark-mass-independent coefficient, $k_1$, has already been given in~\cite{Luscher:2010iy} and reads in our conventions:
\begin{multline}
k_1\of{\lambda/\lambda_0}=\of{2\,\log\of{\lambda/\lambda_0}+\gamma_E}\,\beta_0\\
+N\of{\frac{52}{9}-3\,\log\of{3}}-N_f\of{\frac{4}{9}-\frac{4}{3}\log\of{2}}\ ,\label{eq:k1def}
\end{multline}
with $\beta_0$ being the leading coefficient of the beta function in the massless $\MSb$ scheme.
The quark-mass-dependent coefficient, $l_1$, is the main result from~\cite{Harlander:2016vzb}, and is in our conventions given by
\[
l_1\of{y}=-\frac{4}{3}\,T_R\,N_f\,\Omega_{1q}\sof{-1/\of{2y}^2}\ ,\label{eq:l1def}
\]
where
\begin{multline}
\Omega_{1q}\of{x}=1-\gamma_E+4\,\log\of{2}+\log\of{-x}+\frac{1}{4x}\\
+\frac{1}{8x^2}\int_{0}^{\infty}\dd{z}\,\e^{\frac{z}{4x}}\of{1+z}\of{1-2z}\frac{u\of{z}\log\of{u\of{z}}}{u^2\of{z}-1}
\end{multline}
with $x=-1/\of{2\,\lambda\,m}^2$ and
\[
u\of{z}=\frac{\sqrt{1+1/z}-1}{\sqrt{1+1/z}+1}\ .
\]

To obtain a perturbative expression for the massive gradient flow beta function, we follow the prescription given in~\cite{Harlander:2021esn}. As a first step we define the beta function \refcorr{in terms of} the derivative of Eq.~\eqref{eq:gfrunningcoupling} with respect to \refcorr{$\lambda$}:
\begin{multline}
\beta_{\mathrm{GF}}=-\frac{\dd\uGF}{\dd\log\of{\lambda/\lambda_0}}=-\lambda\frac{\dd\uGF}{\dd\lambda}=\\
-\of{\lambda\frac{\dd k_1\of{\lambda/\lambda_0}}{\dd\lambda}+\lambda\frac{\dd l_1\of{\lambda\,m\of{\lambda_0}}}{\dd\lambda}}\frac{\uMSb^2\of{\lambda_0}}{\of{4\pi}^2}\\
+\order\of{\uMSb^3}\ .\label{eq:gfbetafunctemp}
\end{multline}
Next we note that inverting Eq.~\eqref{eq:gfrunningcoupling} yields
\begin{multline}\label{eq:gfrunningcouplinginv}
\uMSb\of{\lambda_0}=\uGF\of{\lambda}\bof{1-k_1\of{\lambda/\lambda_0}\frac{\uGF\of{\lambda}}{\of{4\pi}^2}\\
-l_1\of{\lambda\,m\of{\lambda_0}}\frac{\uGF\of{\lambda}}{\of{4\pi}^2}+\order\of{\uGF^2\of{\lambda}}}\ ,
\end{multline}
and Eq.~\eqref{eq:gfbetafunctemp} can therefore be written as:
\begin{multline}
\beta_{\mathrm{GF}}\of{\uGF,\lambda\,m}=\\
-\bof{\underbrace{\lambda\frac{\dd k_1\of{\lambda/\lambda_0}}{\dd\lambda}+\lambda\frac{\dd l_1\of{\lambda\,m\of{\lambda_0}}}{\dd\lambda}}_{2\,\beta_{0,\mathrm{GF}}}}\frac{\uGF^2}{\of{4\pi}^2}\\
+\order\of{\uGF^3}\ ,\label{eq:gfbetafunc}
\end{multline}
where $\beta_{0,\mathrm{GF}}$ is the quark-mass-dependent leading coefficient of the gradient flow beta function. After plugging in the definitions from Eqn.~\eqref{eq:k1def} and~\eqref{eq:l1def}, we find that it is purely a function of $\ssof{\lambda\,m}$:
\[
\beta_{0,\mathrm{GF}}\of{x}=\beta_0+\frac{4}{3}\,T_R\,N_f\,x\,\frac{\dd\Omega_{1q}\of{x}}{\dd x}\ ,\label{eq:gfbetafunccoeff0}
\]
where $x=-1/\of{2\,\lambda\,m}^2$, as before.

Note that the beta function coefficient in Eq.~\eqref{eq:gfbetafunccoeff0} reduces in the limits $\ssof{m\to0}$ and $\ssof{m\to\infty}$ to the $\MSb$ coefficients for $N_f$ massless fermions and pure gauge, respectively, as is the case also for the \refcorr{1- and 2-loop} coefficients in the massive BF-MOM scheme~\cite{Rebhan:1985yf,Jegerlehner:1998zg} used in the main text.

\begin{figure}[ht]
\centering
\begin{minipage}[t]{0.85\linewidth}
\includegraphics[width=\linewidth,keepaspectratio]{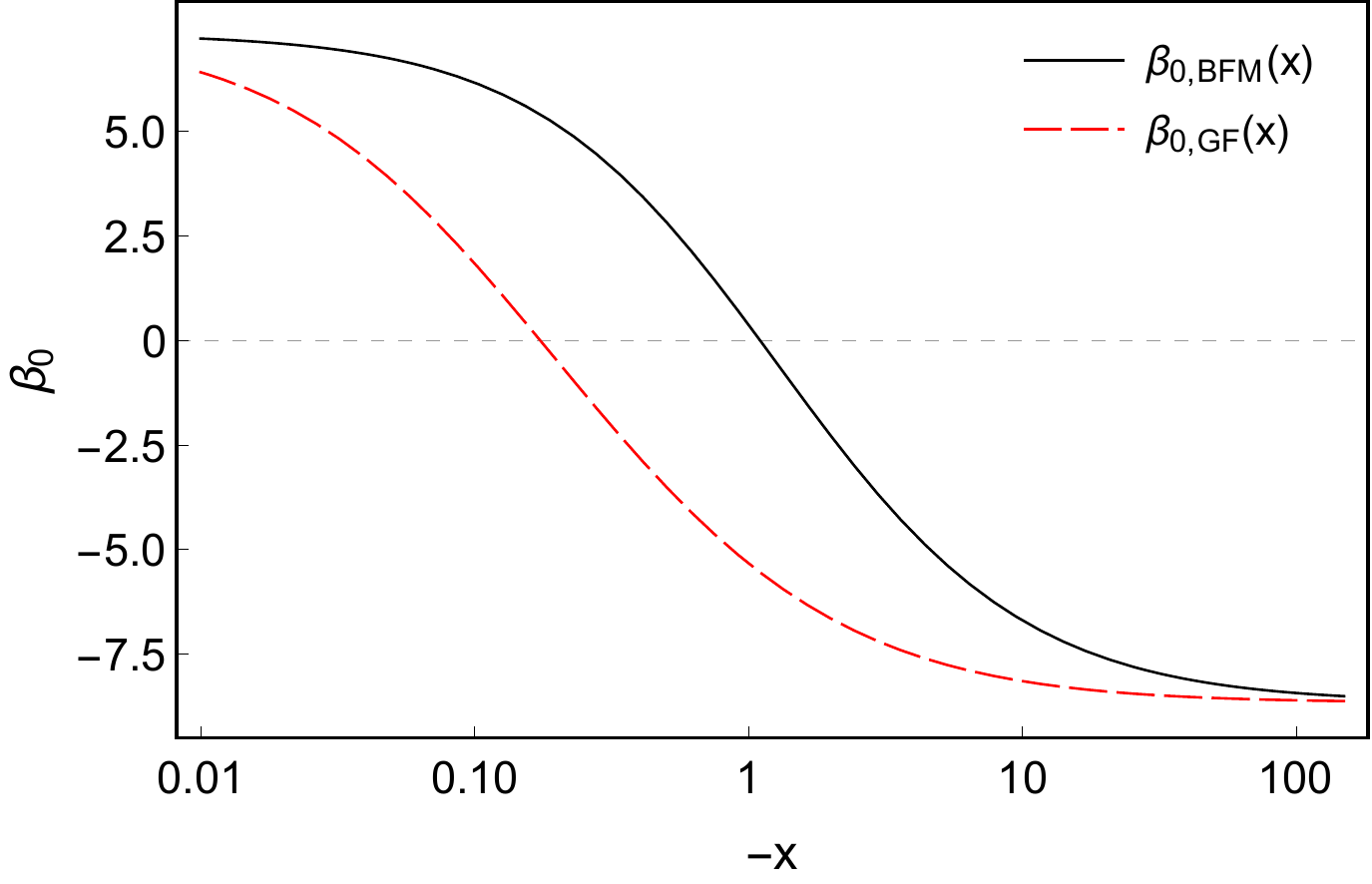}\\[5pt]
\includegraphics[width=\linewidth,keepaspectratio]{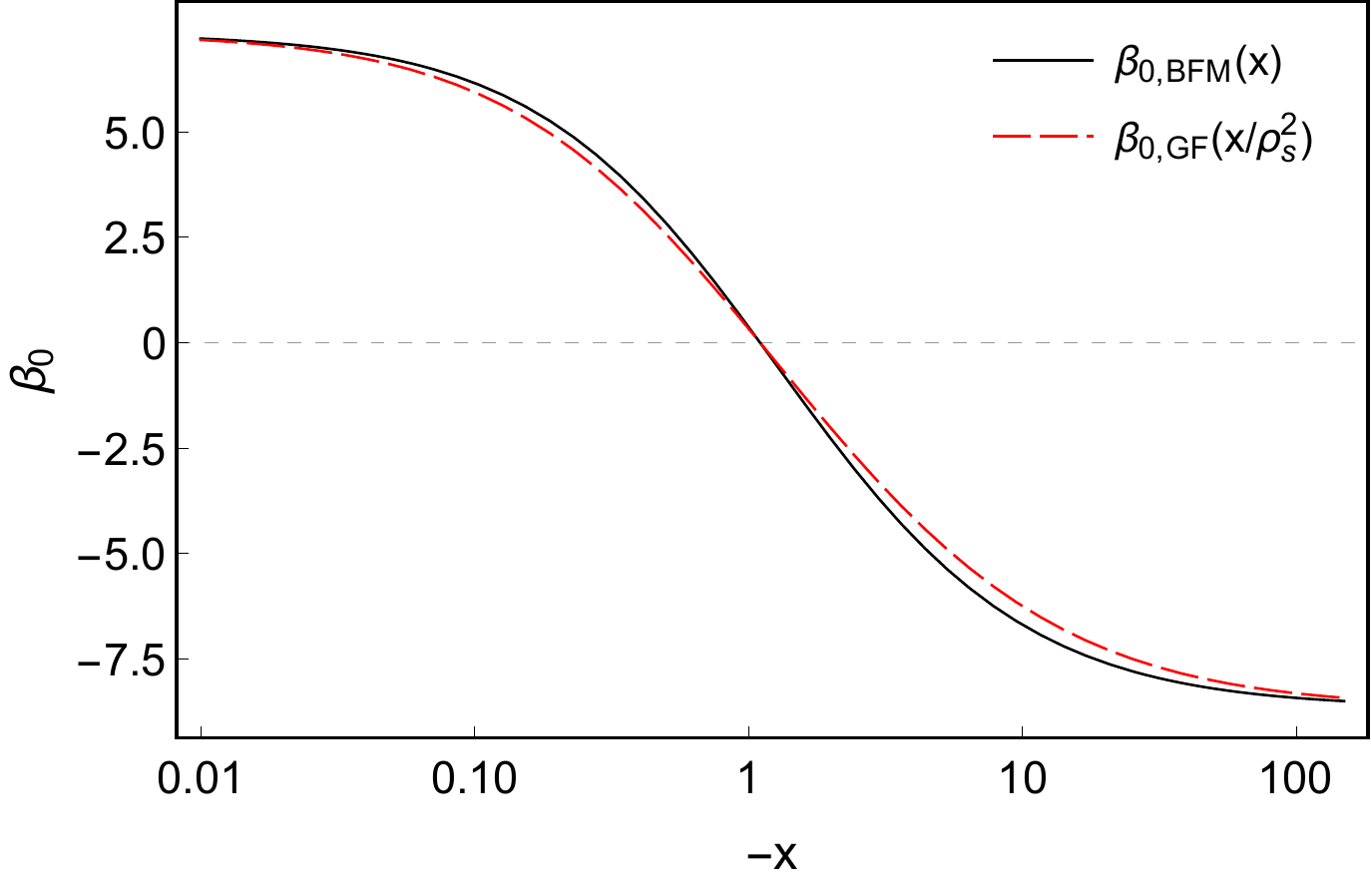}
\end{minipage}
\caption{Comparison of the leading beta-function coefficients in the massive GF (dashed red) and massive BF-MOM (solid black) scheme as functions of $-x=1/(2\,\lambda\,m)^2$. The upper panel shows the situation before and the lower panel after rescaling $x\to x/\rho_s^2$ in the GF scheme.}
\label{fig:betacoeffcomp}
\end{figure}

A few more remarks are in order: {\em firstly}, as the quark-mass-dependency of the gradient flow coupling in Eq.~\eqref{eq:gfrunningcoupling} is known only to leading order, we do not have the necessary information for determining the two-loop coefficient, $\beta_{1,\mathrm{GF}}\of{x}$, of the massive GF-scheme beta function. The same applies for the leading coefficient of the quark mass anomalous dimension\refcorr{, which would also come into play at two-loop order, if the quark mass parameter utilized in~\cite{Harlander:2016vzb} is (as we assumed in Eq.~\eqref{eq:gfrunningcoupling}) indeed meant to be a running mass}. {\em Secondly}, in Eq.~\eqref{eq:gfrunningcoupling} we could set $\lambda_0=\lambda$, but the result Eq.~\eqref{eq:gfbetafunccoeff0} would remain the same. The scale-dependency in $k_1$ is precisely such that it compensates for the case where $\uMSb$ is evaluated at a different scale than $\lambda$. If $\lambda_0=\lambda$, $k_1$ is independent of $\lambda$, but at the same time $\uMSb$ becomes $\lambda$-dependent, so that the $\beta_0$-term in Eq.~\eqref{eq:gfbetafunccoeff0} will be provided by the $\MSb$ beta function instead of by the derivative of $k_1$. Setting $\lambda_0=\lambda$ would also mean that the quark mass would now depend on $\lambda$, but as the quark mass anomalous dimension is to leading order proportional to $\uGF$, the change from $m\of{\lambda_0}$ to $m\of{\lambda}$ in Eq.~\eqref{eq:gfbetafunc} would not affect $\beta_{0,\mathrm{GF}}$, but only terms which are of higher order in $\uGF$. \refcorr{However, with a fully consistent two-loop version of Eq.~\eqref{eq:gfrunningcoupling}, also this ambiguity in the mass-dependency should disappear.}

\paragraph{Relating the GF and the BF-MOM scheme. --}\label{par:gfvsbfm}
In order to relate the massive GF and the BF-MOM scheme for our $N_f=24$ theory, we note that the length scale $\lambda$ in the GF scheme is not properly normalized. This can be seen in the upper panel of Fig.~\ref{fig:betacoeffcomp}, where the leading coefficient of the massive GF beta function is plotted as function of $-x=1/\ssof{2\lambda\,m}^2$, together with the corresponding coefficient of the BF-MOM scheme \refcorr{(as we are at one-loop order, we consider the masses in both schemes as non-running)}. The change of sign in the leading beta function coefficient, which indicates fermion decoupling, does for the GF scheme not occur at the expected decoupling point where $x\approx-1$.

However, as mentioned already below Eq.~\eqref{eq:gfbetafunccoeff0}, in both schemes, the leading beta-function coefficients approach in the limits $\ssof{x\to-\infty}$ and $\ssof{x\to0}$ the same asymptotic values, which are given by the value of the universal leading $\MSb$ coefficient, $\beta_0$, for $N_f=24$ massless fermions, and for pure gauge, respectively. Only the scale at which the transition between the two asymptotic values \refcorr[happens]{takes place} is not as expected for the massive GF scheme.

As the RG equations~\eqref{eq:massivergeq} depend only on relative scale changes, we can adjust the length scale in the GF scheme relative to that of the BF-MOM scheme by requiring $\lGF=\rho_s\lBFM$ for some $\rho_s>0$, and tuning $\rho_s$ till the behavior of the running coupling near the decoupling point becomes in the GF-scheme as similar as possible to that of the BF-MOM scheme, if the same initial conditions are used.

In an asymptotically free theory, such a matching would be carried out in the vicinity of the trivial fixed point, where the running coupling and running mass should approach universal values~\cite{Sint:1998iq}. For our theory with $N_f=24$ massive fermion flavors, which does not possess such a trivial fixed point, a set of points where some sort of universality in the behavior of running coupling and running mass could be expected across schemes, is given by the fermion-mass-dependent decoupling points. At these points, the beta function has to change sign, scheme-independently, and the quark mass acquires a physical meaning as it sets the actual energy scale at which this sign change of the beta function occurs. \refcorr{Note that a similar argument is used to define a pole mass $m_0$ in terms of a running mass $m\of{\mu}$ that depends on a renormalization energy scale $\mu$: the pole mass is defined by requiring $m\of{\mu=m_0}=m_0$~\cite{Hagiwara:1982ct}.}

\begin{figure}[ht]
\centering
\begin{minipage}[t]{0.82\linewidth}
\includegraphics[height=\linewidth,keepaspectratio,right]{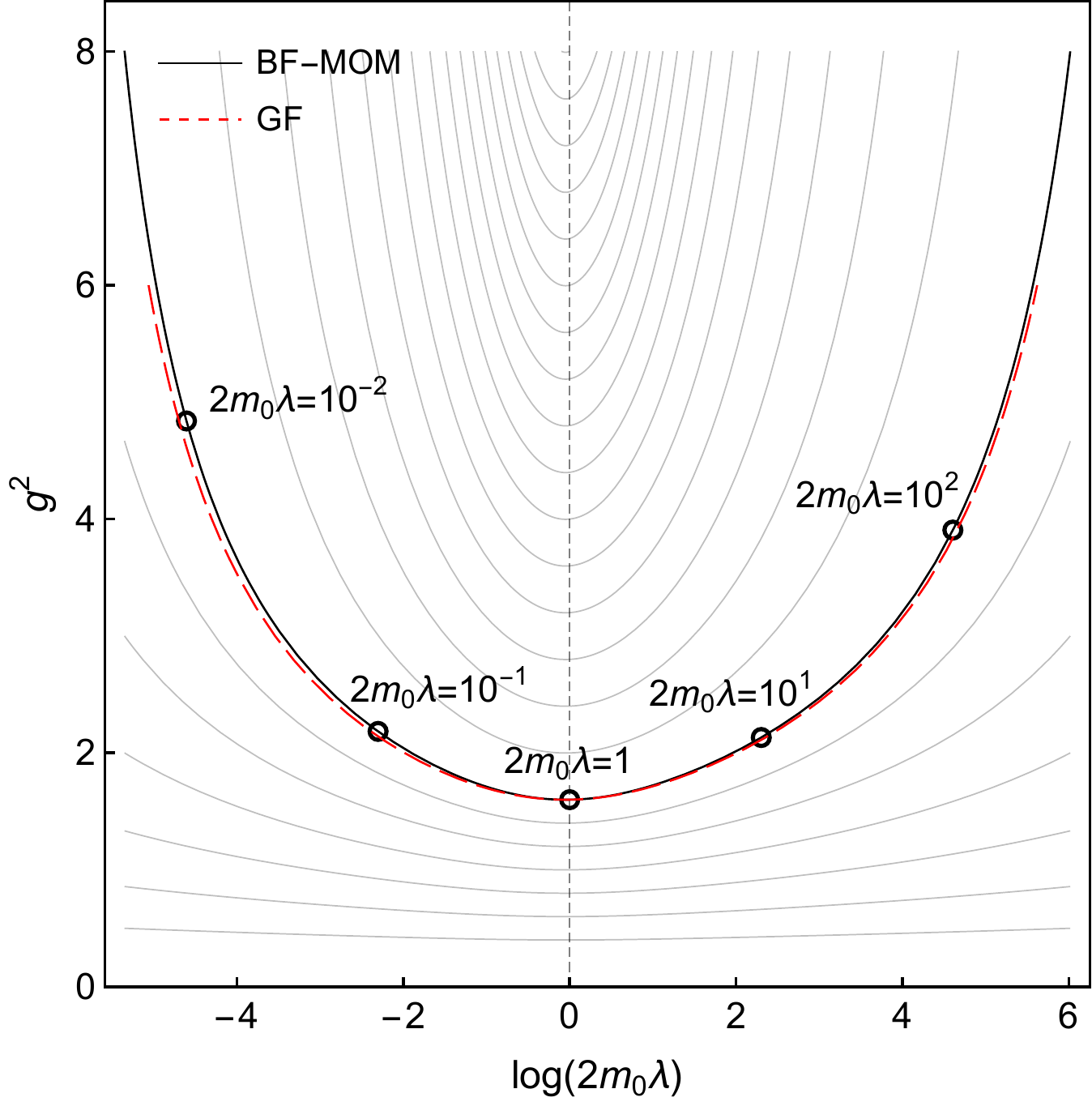}
\end{minipage}\\[-5pt]
\caption{Comparison of the one-loop running couplings in the (rescaled) massive GF (dashed red) and massive BF-MOM (solid black) scheme as functions of the log of flow-scale $\lambda$ relative to the decoupling length scale $\lambda_0=1/\ssof{2\,m_0}$.}
\label{fig:runningcouplingcomp}
\end{figure}

To make the matching procedure more precise, \refcorr{we assume for the moment that the RG equations are in both schemes parametrized in terms the same type of mass-parameter (either pole mass or running mass),} and consider the Taylor expansions of the running coupling and quark mass from Eq.~\eqref{eq:taylorexpandedflow} in both schemes, using:
\begin{subequations}
\begin{align}
\lBFM &= \lambda_{\phantom{0}}\ ,\ \lGF = \rho_s\lambda\ ,\\
\lBFM_0 &= \lambda_0\ ,\ \lGF_0 = \rho_s\lambda_0\ ,\\
\uGF\ssof{\lGF_0} &= \uBFM\ssof{\lBFM_0}=u_0\ ,\\
m_{\mathrm{GF}}\ssof{\lGF_0} &= m_{\mathrm{BFM}}\ssof{\lBFM_0}=m_0\ ,
\end{align}
\end{subequations}
and then set $\lambda_0=1/\ssof{2\,m_0}$, corresponding to the approximate decoupling point in the BF-MOM scheme. Now we require that
\begin{subequations}
\begin{align}\label{eq:equalrunning}
\uGF\ssof{\lGF} &= \uBFM\ssof{\lBFM}\ ,\\
m_{\mathrm{GF}}\ssof{\lGF} &= m_{\mathrm{BFM}}\ssof{\lBFM}\ ,
\end{align}
\end{subequations}
is satisfied as well as possible in at least some neighborhood of $\lambda=\lambda_0$. As the factors of $\rho_s$ cancel in ratios $\lGF/\lGF_0$, the Taylor expansions are in both schemes power series in $\log\ssof{\lambda/\lambda_0}$. However, $\rho_s$ does not cancel in $\lGF_0 m_{\mathrm{GF}}\ssof{\lGF_0}=\rho_s\lambda_0 m_0$. The requirement \eqref{eq:equalrunning} is therefore equivalent to:
\[
c^{\ssof{\mathrm{GF}}}_n\of{u_0,x_0/\rho_s^2}=c^{\ssof{\mathrm{BFM}}}_n\of{u_0,x_0}\ \forall n\in\mathbb{N}\ ,\label{eq:equalrunningreq}
\]
where $x_0=-1/\ssof{2m_0\lambda_0}^2=-1$. Unfortunately, as for the GF scheme, we do only have access to the leading beta function coefficient, we can impose Eq.~\eqref{eq:equalrunningreq} only to leading order in the equation for the running coupling, namely:
\[
\beta_{0,\mathrm{GF}}\of{x_0/\rho_s^2}=\beta_{0,\mathrm{BFM}}\of{x_0}\ ,\label{eq:betacoeffeq}
\]
which yields the rescaling factor
\[
\rho_s=2.535945\ldots\ .\label{eq:rescalefac}
\]
\refcorr{As we are restricted to one-loop order, we can neglect issues related to the definition of mass (running vs. pole-mass) in the two schemes, as the running of the mass would affect the running of the coupling only at higher loop order.}

The lower panel of Fig.~\ref{fig:betacoeffcomp}, shows again a comparison of the leading beta function coefficients of the BF-MOM and the GF scheme, as functions of $-x=1/\ssof{2\,\lambda\,m}^2$, after the rescaling $x\to x/\rho_s^2$ has been applied in the GF scheme. As can be seen, the value of $\rho_s$, obtained by requiring Eq.~\eqref{eq:betacoeffeq} to hold at the approximate decoupling point $x_0=-1$, leads to a relatively good overall agreement between the leading beta-function coefficients of the two schemes. As a consequence, the one-loop running coupling in the GF scheme agrees after the rescaling $x\to x/\rho_s^2$ remarkably well with the one-loop running coupling in the BF-MOM scheme\refcorr{, which is illustrated in Fig.~\ref{fig:runningcouplingcomp}}.
\begin{figure}[ht]
\centering
\begin{minipage}[t]{0.49\linewidth}
\centering
\includegraphics[width=\linewidth,keepaspectratio]{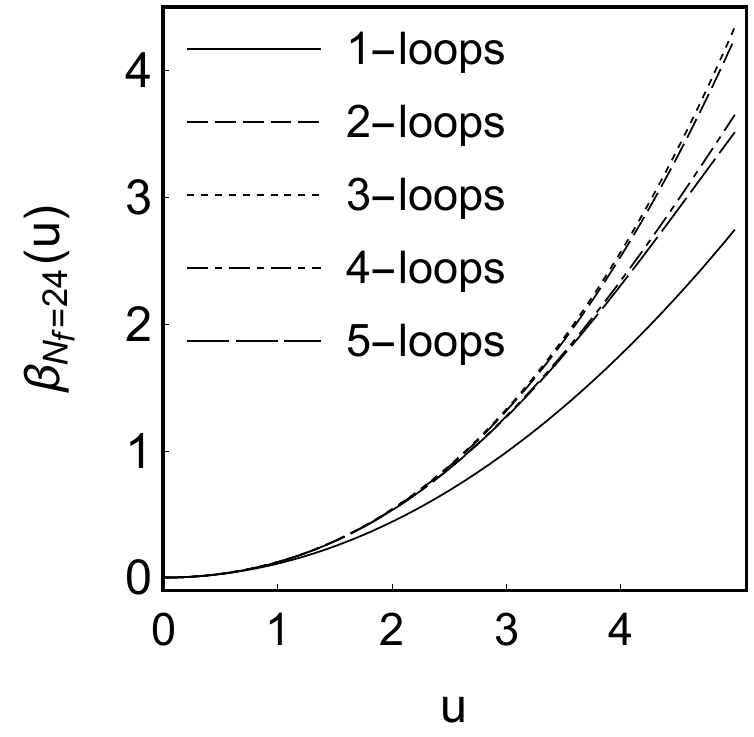}
\end{minipage}\hfill
\begin{minipage}[t]{0.49\linewidth}
\centering
\includegraphics[width=\linewidth,keepaspectratio]{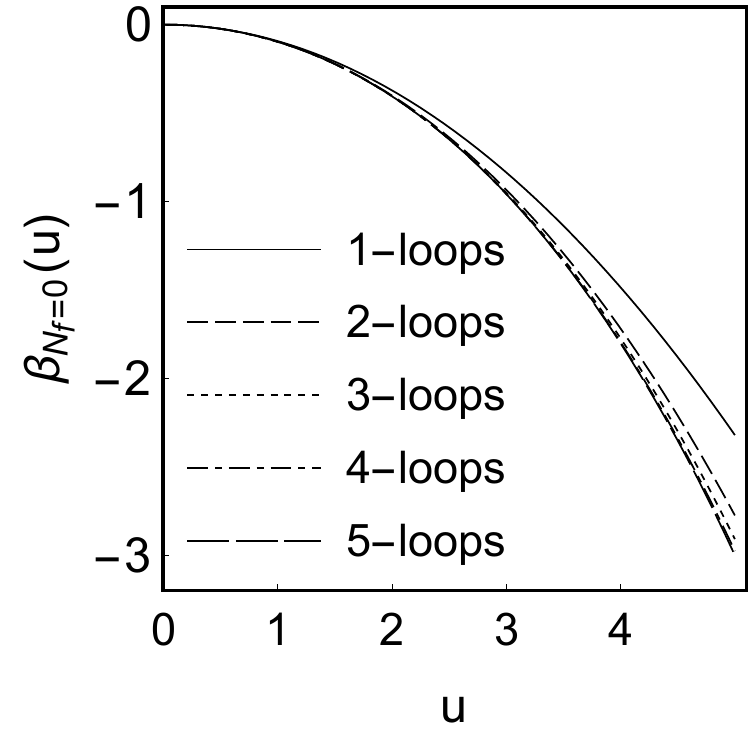}
\end{minipage}\\[-5pt]
\caption{Comparisons of the 1- to 5-loop $\MSb$ beta functions for $\SU{2}$ gauge theory with $N_f=24$ massless flavors (left) and pure gauge, i.e. $N_f=0$ flavors (right).}
\label{fig:betaloopcomp}
\end{figure}

\paragraph{1-loop vs. 2-loop BF-MOM beta functions. --}\label{par:1lvs2lbfmbf}
Ideally, we would compare our lattice data for the non-perturbative running GF coupling with predictions obtained within the perturbative, massive GF scheme \refcorr{discussed above}. However, as we know only the leading 1-loop beta function coefficient from Eq.~\eqref{eq:gfbetafunccoeff0} for the massive GF scheme, the corresponding perturbative 1-loop running coupling cannot be expected to \refcorr[accurately represent]{be a good approximation to} the full, non-perturbative running coupling if $\uGF>2$. This estimate is based on a comparison \refcorr{of the $\MSb$ beta functions for $\SU{2}$ with $N_f=24$ and $N_f=0$ massless flavors at different loop orders (cf. Fig.~\ref{fig:betaloopcomp}). The one- and two-loop massive GF and massive BF-MOM beta functions converge both to these massless $\MSb$ beta functions} in the limiting cases of $\ssof{x\to -\infty}$ and $\ssof{x\to 0}$.
\begin{figure}[ht]
\centering
\begin{minipage}[t]{0.82\linewidth}
\centering
{\small 1-loop}\\
\includegraphics[width=\linewidth,keepaspectratio]{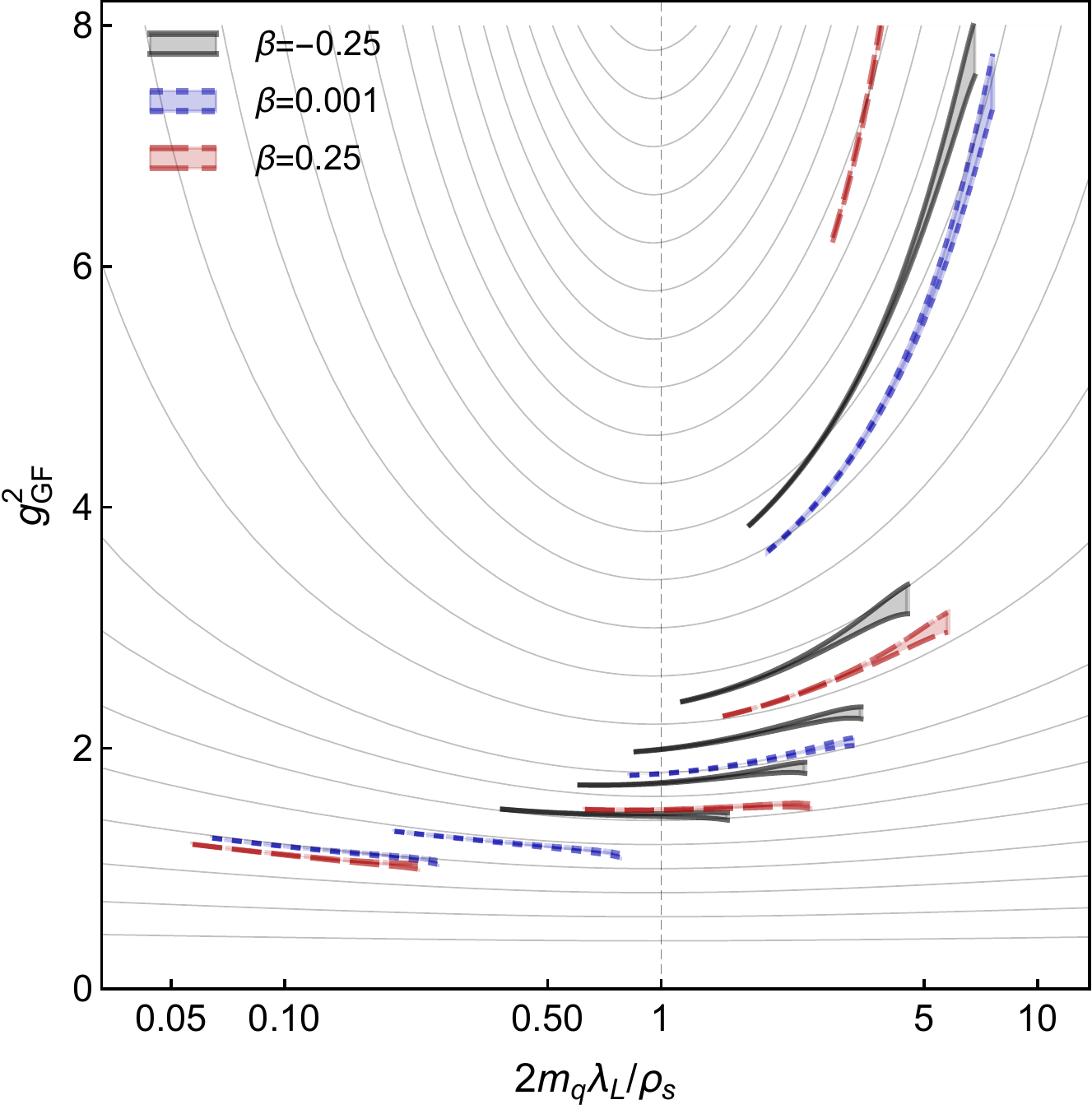}
\end{minipage}\\[7pt]
\begin{minipage}[t]{0.82\linewidth}
\centering
{\small 2-loop}\\
\includegraphics[width=\linewidth,keepaspectratio]{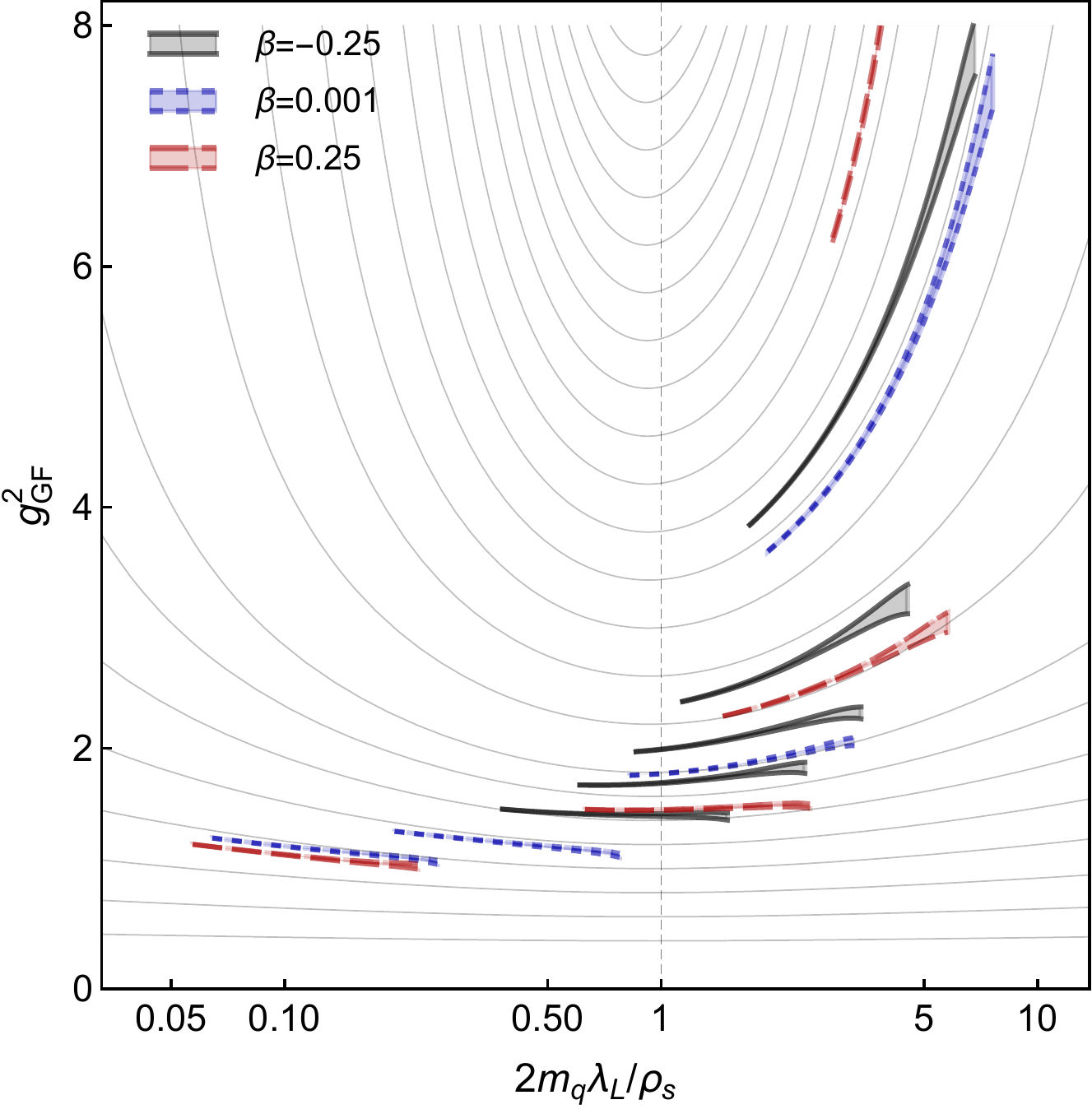}
\end{minipage}\\[-5pt]
\caption{The two panels show the lattice results for the GF running coupling (black, blue and red error bands) as functions of $2\,m_q\,\lambda_L/\rho_s$, with $m_q$ being the PCAC quark mass and $\rho_s$ as in Eq.~\ref{eq:rescalefac}. The data is superimposed to a family of 1-loop (upper panel) and 2-loop (lower panel) BF-MOM running coupling curves.}
\label{fig:runningcouplingloopcomp}
\end{figure}

From our simulation data\refcorr{, only a small number of running coupling trajectories fall within this limited range and do not cover the heavy quark case}. With the 2-loop running coupling, one could hope to find good agreement with the simulation data for $\uGF$ up to approximately \refcorr{$\uGF\sim 3$}. This would require knowledge of at least the 2-loop beta function coefficient and, if the mass-dependency of the RG functions is parametrized in terms of a running mass, also of the 1-loop gamma function coefficient, which is currently not available for the massive GF scheme. We therefore compare our data directly with the 2-loop running coupling in the BF-MOM scheme, after having converted the lattice $\lambda_L\,m_q$, which is a GF scheme quantity, to the BF-MOM scheme, by dividing by the factor $\rho_s$ from Eq.~\eqref{eq:rescalefac}.

Fig.~\ref{fig:runningcouplingloopcomp} illustrates the discrepancies in the agreement between the 1- and 2-loop BF-MOM running coupling and the scale-matched lattice data. As can be seen, the 2-loop running coupling captures the behavior of the lattice data much better. As the perturbative 1-loop BF-MOM and 1-loop massive GF running coupling after scale matching show in Fig.~\ref{fig:runningcouplingcomp} very good agreement, we expect that Fig.~\ref{fig:runningcouplingloopcomp} would look very similar if one could replace the 1- and 2-loop BF-MOM running couplings with the corresponding 1- and 2-loop massive GF running coupling.

Note that our data for PCAC quark masses is unrenormalized. The missing multiplicative renormalization introduces small relative x-axis offsets (in log scale) between the different lattice running coupling curves in Fig~\ref{fig:runningcouplingloopcomp}. However, this does not significantly affect the overall \refcorr{qualitative} agreement between the lattice data and the 2-loop perturbative running coupling curves.

\paragraph{3-loop non-universality. --}\label{par:3lnonuniv}
\refcorr{In the previous section we argued that up to 2-loop order the beta functions of both, the massive GF and massive BF-MOM scheme, approach corresponding $\MSb$ beta functions for either $N_f=24$ massless fermions if the renormalization length scale is much smaller than the inverse quark mass ($\lambda\ll 1/m$), and for pure gauge ($N_f=0$) if the renormalization length scale is much larger than the inverse quark mass ($\lambda\gg 1/m$). This is due to universality of the two-loop beta function among (a class of) massless renormalization schemes.}
\begin{figure}[ht]
\centering
\begin{minipage}[t]{0.48\linewidth}
\centering
\includegraphics[height=0.98\linewidth,keepaspectratio,right]{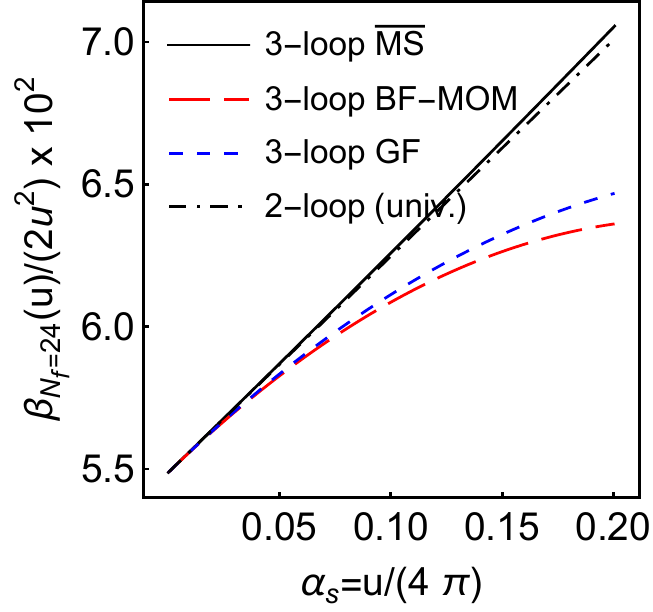}\\[7pt]
\includegraphics[height=0.98\linewidth,keepaspectratio,right]{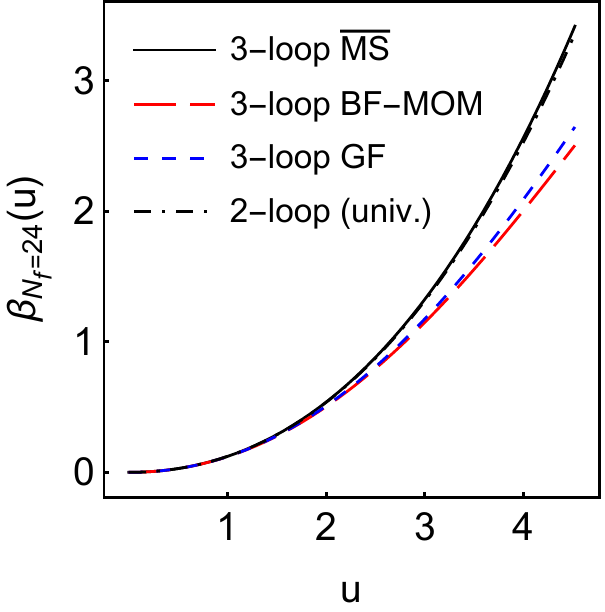}
\end{minipage}\hfill
\begin{minipage}[t]{0.48\linewidth}
\centering
\includegraphics[height=0.98\linewidth,keepaspectratio,right]{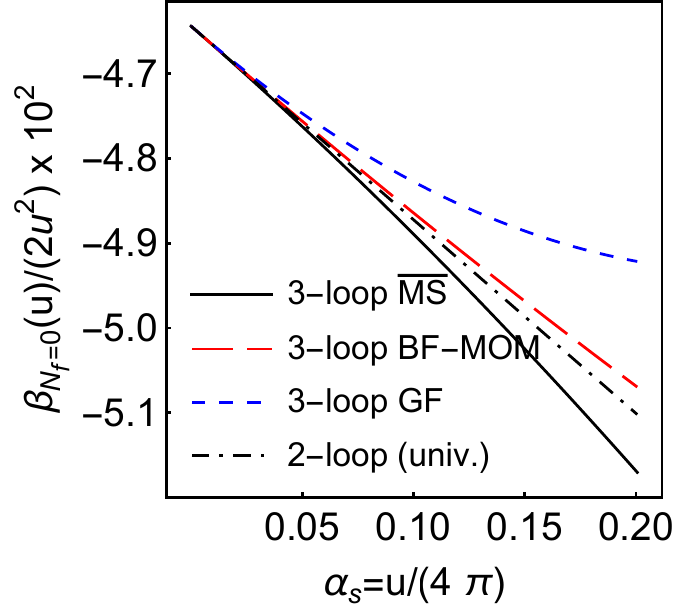}\\[7pt]
\includegraphics[height=0.98\linewidth,keepaspectratio,right]{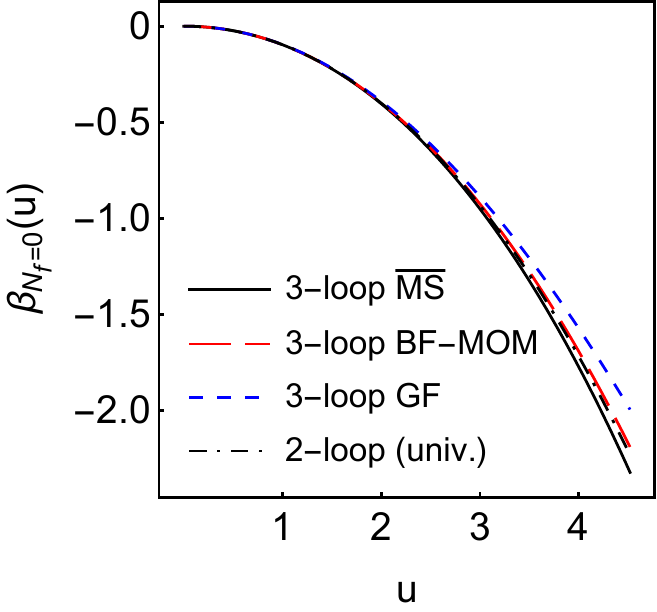}
\end{minipage}\\[-5pt]
\caption{\refcorr{Comparisons of the 3-loop beta function for $\SU{2}$ gauge theory with $N_f=24$ massless flavors (left) and $N_f=0$ flavors (right) in the $\MSb$ (solid, black), BF-MOM (long dashes, red) and GF scheme (short dashes, blue). For comparison also the corresponding 2-loop beta function is shown (dot-dashed, black).}}
\label{fig:betaschemecomp}
\end{figure}

At 3-loop order, however, this universality is in general lost. In~\cite{DallaBrida:2019wur} the discrepancy between the 3-loop beta functions for $\SU{3}$ pure gauge theory in the $\MSb$ and GF scheme has been investigated and it has been shown that the deviation is stronger than one might expect from simply looking at the convergence of different loop orders within one scheme.

We repeat here this analysis for $\SU{2}$ with $N_f=24$ and $N_f=0$  massless fermions, and compare the asymptotic 3-loop beta functions in the GF and BF-MOM scheme.

The starting point is an expression for the running coupling in a massless scheme R in terms of a power series expansion in the $\MSb$ running coupling. The expansion coefficients can for R=GF be found in~\cite{Harlander:2016vzb} and for R=BF-MOM in \cite{Jegerlehner:1998zg}. In contrast to the discussion following Eq.~\eqref{eq:gfrunningcoupling}, we set this time $\lambda_0=\lambda$, so that the expansion coefficients $k_{i,\mathrm{R}}$ become pure numbers:
\begin{multline}\label{eq:gfrunningcouplingmassless}
\uR\of{\lambda}=\uMSb\of{\lambda}\bof{1+k_{1,\mathrm{R}}\,\frac{\uMSb\of{\lambda}}{\of{4\pi}^2}\\
+k_{2,\mathrm{R}}\,\frac{\uMSb^2\of{\lambda}}{\of{4\pi}^4}+\order\of{\uMSb^3\of{\lambda}}}\ .
\end{multline}
Then we proceed as with Eq.~\eqref{eq:gfrunningcoupling}, to find the first three beta function coefficients of the massless scheme R in terms of the known $\MSb$ coefficients:
\begin{subequations}
\[
\beta_{0,\mathrm{R}}=\beta_{0,\MSb}\ ,\quad \beta_{1,\mathrm{R}}=\beta_{1,\MSb}\ ,
\]
and
\[
\beta_{2,\mathrm{R}}=\beta_{2,\MSb}-k_{1,\mathrm{R}}\,\beta_{1,\MSb}+\of{k_{2,\mathrm{R}}-k_{1,\mathrm{R}}^2}\,\beta_{0,\MSb}\ .
\]
\end{subequations}
Note, that here two scheme-matching coefficients, $k_{1,\mathrm{R}}$ and $k_{2,\mathrm{R}}$, are sufficient to obtain the beta function coefficients for scheme R up to 3-loop order in terms of known $\MSb$ coefficients~\cite{Larin:1993tp}. This is only possible because $R$ is here considered to be a massless scheme, so that $k_{1,\mathrm{R}}$ and $k_{2,\mathrm{R}}$ do not depend on any scale; if R were a massive scheme, then also $k_{3,\mathrm{R}}$ would be needed to obtain the 3-loop beta function coefficient in the R-scheme consistently with this method.
\begin{figure}[ht]
\centering
\begin{minipage}[t]{0.82\linewidth}
\centering
{\small \quad asymptotic 3-loop GF}\\
\includegraphics[width=\linewidth,keepaspectratio]{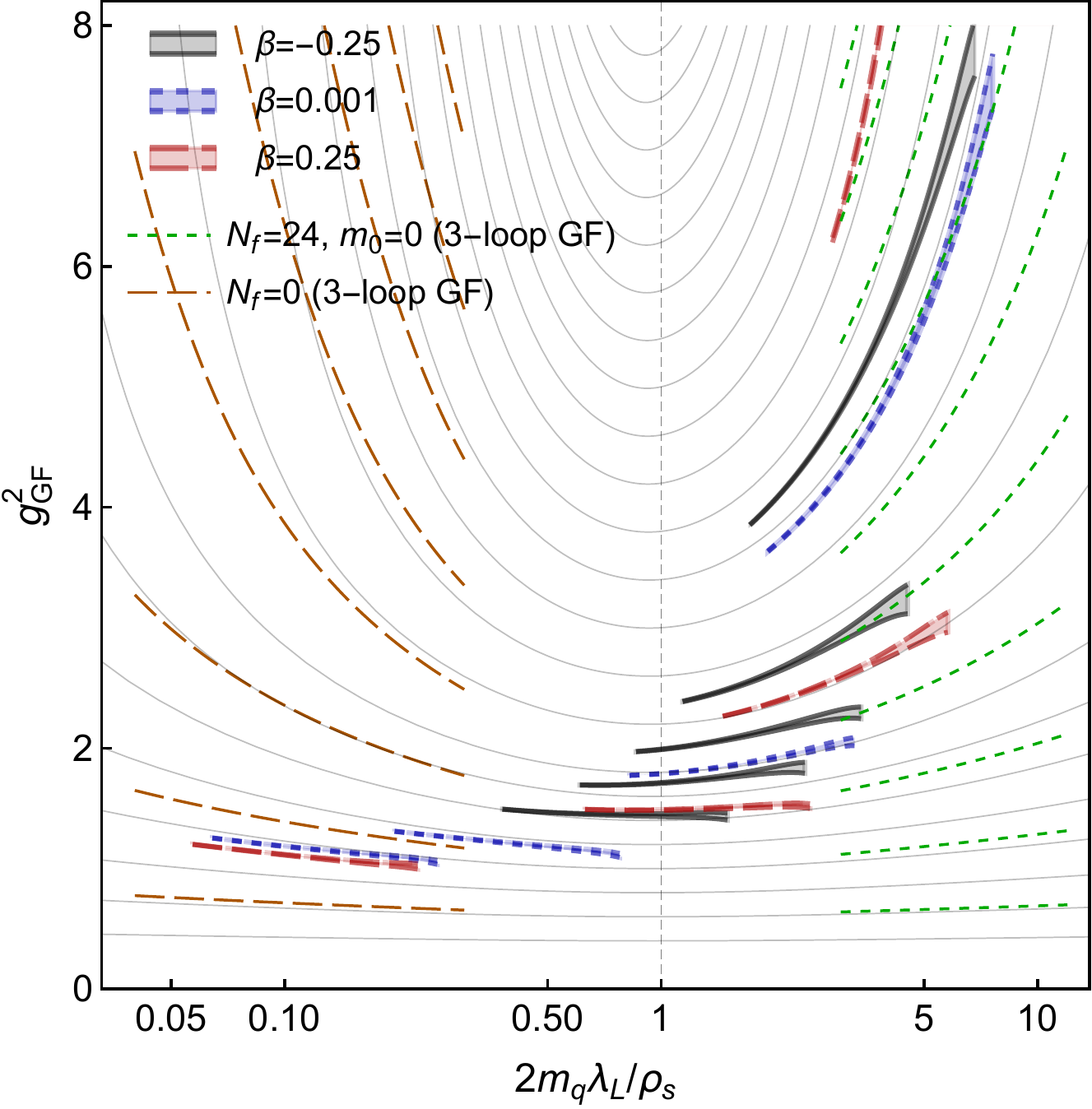}
\end{minipage}\\[7pt]
\begin{minipage}[t]{0.82\linewidth}
\centering
{\small \quad asymptotic 3-loop BF-MOM}\\
\includegraphics[width=\linewidth,keepaspectratio]{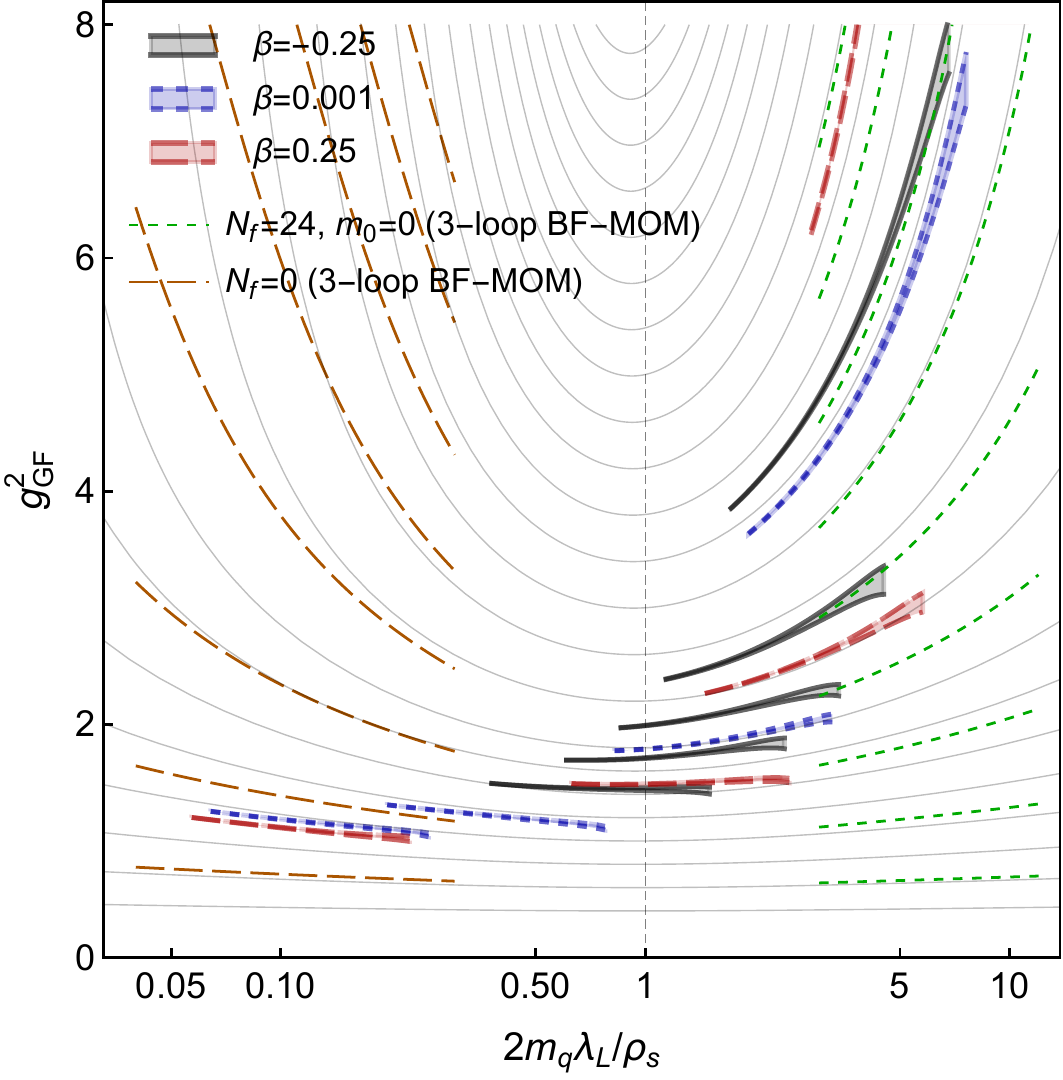}
\end{minipage}\\[-5pt]
\caption{\refcorr{The two panels show the data for the 2-loop massive BF-MOM running coupling (solid, gray lines) and the non-perturbative GF running coupling data (black, blue and red error bands) as in the lower panel of Fig.~\ref{fig:runningcouplingloopcomp}. This is superimposed with running coupling data for the asymptotic casses of $N_f=24$ massless fermions (long dashed, red) and pure gauge (short dashed, green), obtained with the massless 3-loop GF (top) and massless BF-MOM (bottom) beta functions, respectively.}}\vskip-10pt
\label{fig:runningcouplingmassless3loopcomp}
\end{figure}

For R=GF, the coefficient $k_{1,\mathrm{GF}}$ is the same as in Eq.~\eqref{eq:k1def} for $\lambda_0=\lambda$ and $k_{2,\mathrm{GF}}$ can be found in~\cite{Harlander:2016vzb}. In our conventions, they can be expressed as:
\begin{subequations}
\begin{multline}
k_{1,\mathrm{GF}}=8\,\of{4\,\pi}\,\sof{0.045741114\,C_G\\
+0.001888798\,T_R\,N_f}
\end{multline}
and
\begin{multline}
k_{2,\mathrm{GF}}=8\,\of{4\,\pi}^2\sof{-0.0136423\,C_G^2\\
+\of{0.006440134\,C_R-0.0086884\,C_G}\,T_R\,N_f\\
+0.000936117\,T_R^2\,N_f^2}\ ,
\end{multline}
\end{subequations}
where for the $\SU{N}$ gauge group and fundamental fermions,
we have $C_G=N$, $C_R=\ssof{N^2-1}/\ssof{2\,N}$, and $T_R=1/2$.

For R=BF-MOM, the k-coefficients from \cite{Jegerlehner:1998zg} are given by:
\begin{subequations}
\[
k_{1,\mathrm{BFM}}=\frac{205}{36}\,C_G-\frac{20}{9}\,T_R\,N_f
\]
and
\begin{multline}
k_{2,\mathrm{BFM}}=\of{\frac{2687}{72}-\frac{57}{8}\zeta_3}\,C_G^2\\
-\of{\frac{158}{9}+8\,\zeta_3}\,C_G\,T_R\,N_f\\
-\of{\frac{55}{3}-16\,\zeta_3}\,C_R\,T_R\,N_f\ .
\end{multline}
\end{subequations}

For completeness, we also give here the $\MSb$ beta function coefficients~\cite{Larin:1993tp}:
\begin{subequations}
\[
\beta_{0,\MSb}=\frac{11}{3}\,C_G-\frac{4}{3}\,T_R\,N_f\ ,
\]
\[
\beta_{1,\MSb}=\frac{34}{3}\,C_G^2-\frac{20}{3}\,C_A\,T_R\,N_f-4\,C_R\,T_R\,N_f\ ,
\]
and
\begin{multline}
\beta_{2,\MSb}=\frac{2857}{54}\,C_G^3+2\,C_R^2\,T_R\,N_f\\
-\frac{205}{9}\,C_R\,C_R\,T_R\,N_f-\frac{1415}{27}\,C_G^2\,T_R\,N_f\\
+\frac{44}{9}\,C_R\,T_R^2\,N_f^2+\frac{158}{27}\,C_G\,T_R^2\,N_f^2\ .
\end{multline}
\end{subequations}

In Fig.~\ref{fig:betaschemecomp} we show a comparison of the 3-loop beta functions for $\SU{2}$ with $N_f=24$ (left panels) and $N_f=0$ (right panels) massless fermions in the BF-MOM and the GF scheme. For comparison, also the corresponding universal 2-loop beta function and the 3-loop $\MSb$ beta function are shown. As can be seen, the massless 3-loop beta functions for the BF-MOM scheme are in general much closer to the corresponding 3-loop GF-scheme beta functions than the $\MSb$-ones.

For $N_f=24$ the 3-loop BF-MOM and GF scheme beta functions start to deviate relatively strongly from the 2-loop BF-MOM beta function if $u>2$. However, as we do not have lattice data for light quarks at $u>2$, the 2-loop BF-MOM beta function should nevertheless show good qualitative agreement with the lattice data (where the latter is available).

For $N_f=0$ the discrepancy between the beta functions in the different schemes is smaller than for the $N_f=24$ case. Again the 3-loop BF-MOM beta function is a better approximation to the 3-loop GF beta function than the 3-loop $\MSb$ beta function would be, and the 2-loop BF-MOM beta function appears to be a good approximation to the 3-loop GF beta function up to $u\sim 3$.

In Fig.~\ref{fig:runningcouplingmassless3loopcomp} we show once more the data from the lower panel of Fig.~\ref{fig:runningcouplingloopcomp}, superimposed with 3-loop running coupling curves for the asymptotic cases of massless $N_f=24$ fermions (long-dashed, red) if $\lambda\ll 1/2\,m_0$ and pure gauge (short-dashed, green) if $\lambda\gg 1/2\,m_0$ in the GF scheme (top) and in the BF-MOM scheme (bottom). If we had access to the 3-loop coefficient for the massive GF and BF-MOM schemes, the corresponding massive running couplings would smoothly interpolate between the asymptotic massless $N_f=24$ and corresponding $N_f=0$ running couplings.
As can be seen, the lattice data (black, blue and red error bands) shows excellent qualitative agreement with the asymptotic, massless 3-loop GF and BF-MOM running couplings, where a comparison is possible.

\end{document}